\documentclass[a4paper,11pt]{article}
\pdfoutput=1 % if your are submitting a pdflatex (i.e. if you have
\usepackage{jheppub} % for details on the use of the package, please
\usepackage[T1]{fontenc} % if needed
\usepackage{bm,bbm}
\usepackage{graphics}
\usepackage{amssymb}
\usepackage{hyperref}
\usepackage{slashed}
\usepackage{mathrsfs}
\usepackage{esint}
\usepackage[normalem]{ulem}
\usepackage{longtable}
\usepackage{cancel}

\newcommand{\msb}{{\overline{\mathrm{MS}}}}

\newcommand{\nn}{\nonumber}
\newcommand{\sfrac}[2]{\mbox{$\frac{#1}{#2}$}}
\definecolor{purple}{rgb}{0.57, 0.36, 0.51}

\title{\boldmath
Reconciling the Contour-Improved and Fixed-Order Approaches for $\tau$ Hadronic Spectral
	Moments II: \\ Renormalon Norm and Application in $\alpha_s$ Determinations
}

\preprint{
\begin{flushright}
UWThPh-2022-7\\
\end{flushright}
}

\author[a]{Miguel A. Benitez-Rathgeb}
\author[a,b]{Diogo Boito}
\author[a,c]{Andr\'e H. Hoang}
\author[a,d]{Matthias Jamin}

\affiliation[a]{ Faculty of Physics, University of Vienna,\\Boltzmanngasse 5, A-1090 Wien, Austria}
\affiliation[b]{Instituto de F\'isica de S\~ao Carlos, Universidade de S\~ao Paulo, \\ CP 369, 13560-970, S\~ao Carlos, SP, Brazil}
\affiliation[c]{Erwin Schr\"odinger International Institute for Mathematics and Physics,\\
University of Vienna, Boltzmanngasse 9, A-1090 Wien, Austria}
\affiliation[d]{Department of Addictive Behaviour, Central Institute of Mental Health, Medical Faculty Mannheim, Heidelberg University, Mannheim, Germany}

\emailAdd{miguel.angel.benitez-rathgeb@univie.ac.at}
\emailAdd{boito@ifsc.usp.br}
\emailAdd{andre.hoang@univie.ac.at}
\emailAdd{matthias.jamin@gmail.com}

\abstract{
In a previous article, we have shown that the discrepancy between the fixed-order (FOPT) and contour-improved (CIPT) perturbative expansions for $\tau$ hadronic spectral function moments, which had affected the precision of $\alpha_s$ determinations for many years, may be reconciled by employing a renormalon-free (RF) scheme for the gluon condensate (GC) matrix element. In addition, the perturbative convergence of spectral function moments with a sizeable GC correction can be improved. The RF GC scheme depends on an IR factorization scale $R$ and the normalization $N_g$ of the GC renormalon. In the present work, we use three different methods to determine $N_g$, yielding a result with an uncertainty of $40\%$. Following two recent state-of-the-art strong coupling determination analyses at ${\cal O}(\alpha_s^5)$, we show that using the renormalon-free GC scheme successfully reconciles the results for $\alpha_s({m_\tau^2})$ based on CIPT and FOPT. The uncertainties due to variations of $R$ and the uncertainty of $N_g$ only lead to a small or moderate increase of the final uncertainty of $\alpha_s(m_\tau^2)$, and affect mainly the CIPT expansion method. The FOPT and CIPT results obtained in the RF GC scheme may be consistently averaged.
The RF GC scheme thus constitutes a powerful new ingredient for future analyses of $\tau$ hadronic spectral function moments.
}

\begin{document}
\maketitle
\flushbottom

\section{Introduction}
\label{sec:intro}

The determination of the QCD strong coupling $\alpha_s$ from the inclusive hadronic $\tau$ decay spectral functions, obtained through the analysis of finite energy sum rules (FESRs) for spectral function moments, represents one of the most precise methods to extract this fundamental quantity of quantum chromodynamics (QCD) from experimental data~\cite{ParticleDataGroup:2020ssz}. For a long time, however, a systematic theoretical uncertainty associated with the renormalization scale setting has persisted. The values of $\alpha_s$ obtained with a
strict fixed-order expansion, which goes under the name fixed-order perturbation theory (FOPT) (see Ref.~\cite{Beneke:2008ad}),  have been systematically lower than the values obtained using the so-called contour-improved perturbation theory (CIPT)~\cite{Pivovarov:1991rh,LeDiberder:1992jjr}, which resums  a certain set of logarithmic phase corrections with the use of the QCD $\beta$-function.

In a recent paper~\cite{Benitez-Rathgeb:2022yqb}, to which we will refer often as `Part I',  we laid the foundations of a method to eliminate the CIPT-FOPT discrepancy.  The method was motiviated from the work of Hoang and Regner~\cite{Hoang:2020mkw,Hoang:2021nlz} where it was shown that the discrepancy can be of infra-red (IR) origin and caused by contributions in the CIPT expansion terms that are incompatible with the standard form commonly adopted for the higher dimensional corrections in the operator product expansion (OPE). Any IR renormalon contributes a certain amount to the discrepancy, which was termed `asymptotic separation'. But numerically the asymptotic separation is strongly dominated by the renormalon associated with the gluon condensate (GC) contribution in the operator product expansion (OPE). The basic idea of the method proposed in Ref.~\cite{Benitez-Rathgeb:2022yqb}, consists in a redefinition of the GC matrix element in order to remove (or subtract) the associated renormalon from the perturbative expansion of the Adler function. This redefinition was called the renormalon-free gluon condensate (RF GC) scheme  and involves perturbative subtractions that allow for a straightforward and transparent implementation in the context of the computations involved for the FOPT and CIPT expansions of the spectral function moments. This perturbative and observable-independent subtraction approach is very common in many areas of particle physics phenomenology where the removal of renormalon divergences are important, most prominently in the context of using short-distance mass schemes in heavy-quark physics~\cite{Hoang:2020iah,Beneke:2021lkq}. In Ref.~\cite{Benitez-Rathgeb:2022yqb} such a subtraction scheme was for the first time applied to the $\tau$ hadronic spectral function moment. The essential conceptual aspect of this seemingly trivial scheme change, which reaches beyond the well established use of short-distance quark masses, is that it substantially modifies the CIPT series coefficients even for spectral function moments where the GC OPE correction is strongly suppressed or even vanishes.\footnote{We call such spectral function moments gluon condensate suppressed (GCS).
For GCS spectral function moments,  when switching to the RF GC scheme, the changes of the FOPT series terms are tiny and comparable to the GC OPE corrections.}
This fact signifies the inconsistency of the CIPT expansion when the $\overline{\rm MS}$ GC scheme is employed~\cite{Hoang:2020mkw,Hoang:2021nlz}. The work of Hoang and Regner laid out the conceptual basis to understand this paradoxical behavior by showing that the Borel representation of the CIPT expansion has a structure that is incompatible with the standard Borel calculus.\footnote{Based on plausibility arguments in the context of Borel model studies, it was stated already in Refs.~\cite{Beneke:2008ad,Beneke:2012vb} that the CIPT expansion is disfavored. However the inconsistency of the CIPT expansion with the standard form of the OPE in the presence of IR renormalons was not known prior to the work by Hoang and Regner.}

A particularly useful property of the RF GC  scheme we have proposed in Part I is that it is scale-invariant, but at the same time has a freely adaptable IR factorization scale $R$,  which should be chosen to be close to the dynamical scale of the observable. This freedom of scale choice is controlled by a renormalization group evolution (called `R-evolution') that is known exactly to all orders and makes the RF GC scheme also suitable  for other processes where the gluon condensate contributes. The implementation of the scheme depends on knowledge about the GC renormalon norm, $N_g$, which must be supplemented independently.

In Part I, we have demonstrated that in the RF GC scheme with the correct value of $N_g$, the CIPT-FOPT discrepancy  can indeed be resolved  and, moreover, is already significantly reduced at  ${\cal O}(\alpha_s^4)$ and ${\cal O}(\alpha_s^5)$, which are the perturbative orders for which QCD corrections are available. While the CIPT-FOPT discrepancy and its reduction in our RF GC scheme take place for GCS spectral function moments, the new scheme is also capable of significantly improving the perturbative behaviour of moments for which the GC OPE correction is sizeable, i.e.\ for so-called GC enhanced (GCE) moments.  In Part~I, this was shown in specific realizations of the perturbative series at higher orders, namely the large-$\beta_0$ limit and using  all-order renormalon models for the Adler function,  where the gluon condensate renormalon norm $N_g$ was exactly known. An application of the RF GC scheme in the context of a fully consistent phenomenological determination of the strong coupling  and a study on the impact of the uncertainty of $N_g$ was missing.

In the present work, which is Part II in this series of papers, we address  these two issues: We carry out a thorough analysis to determine the GC renormalon norm $N_g$ using three different methods, and we conduct consistent phenomenological determinations of $\alpha_s$ following two state-of-the-art approaches from the recent literature~\cite{Pich:2016bdg,Boito:2020xli}, accounting for the uncertainty of the norm $N_g$ and variations of the IR factorization scale $R$. We emphasize that the main aim of our $\alpha_s$ determinations is to demonstrate the effectiveness of the RF GC scheme to resolve the CIPT-FOPT discrepancy in the context of accounting for all perturbative uncertainties in a realistic manner. In particular, our analyses are not intended to provide a discussion on the merits of the different ways the nonperturbative corrections in Refs.~\cite{Pich:2016bdg,Boito:2020xli} have been accounted for. We emphasize that our two analyses are not identical copies of the complete set of analyses carried out in Refs.~\cite{Pich:2016bdg,Boito:2020xli} as this would go way beyond the scope of this article. The main aim of them is to demonstrate the practical use of the RF GC scheme in achieving precise strong coupling determinations that are not affected by the CIPT-FOPT discrepancy problem. The essential message is that in the RF GC scheme the CIPT results closely approach the FOPT results and that the additional sources of theoretical uncertainties in the RF GC scheme (related to $R$ and $N_g$) are relatively small.

The basis of our analysis is the proposition that the GC renormalon gives a sizeable contribution to the known QCD corrections of the Adler function at ${\cal O}(\alpha_s^4)$ (including the state-of-the-art estimates for the ${\cal O}(\alpha_s^5)$ coefficient~\cite{Jamin:2021qxb,Boito:2018rwt,Caprini:2019kwp,Baikov:2008jh}), such that $N_g$ is sizeable as well and can be determined with reliable uncertainties from the ${\cal O}(\alpha_s^4)$ (or ${\cal O}(\alpha_s^5)$) perturbative coefficients. This proposition has been advocated as being natural in Ref.~\cite{Beneke:2008ad}. We have furthermore shown in Part~I through a dedicated analysis, that the perturbative behavior of the spectral function moments is fully consistent with this proposition. It should be mentioned, however, that it is in principle not excluded that this behavior may be mimicked at ${\cal O}(\alpha_s^4)$ through an interplay of other renormalons and sizeable lower-order convergent QCD corrections in the presence of an extremely small (and practically negligible) value of $N_g$~\cite{DescotesGenon:2010cr}. However, as shown in Ref.~\cite{Beneke:2012vb} such a scenario would correspond to an unnatural and implausible fine-tuned renormalon structure of the Adler function. More light could potentially be shed on this matter through the actual calculation of the QCD corrections at ${\cal O}(\alpha_s^5)$ and beyond, which, however, may appear unlikely in the near future. Our analysis therefore relies on the proposition that the observed consistency of the known QCD corrections with a sizeable GC norm $N_g$ is not caused by some fine-tuned renormalon scenario with a negligible value for $N_g$ or some purely accidental behavior of the ${\cal O}(\alpha_s^4)$ QCD corrections. The same proposition is in principle also applied when short-distance quark mass renormalization schemes employed in order to achieve an improved perturbative behavior for heavy quark mass sensitive observables~\cite{Hoang:2020iah,Beneke:2021lkq}.

We note that, the RF GC scheme is formulated using the $C$-scheme for the strong coupling~\cite{Boito:2016pwf} for the concrete value $C=0$, see  Sec.~2.2 of Part I~\cite{Benitez-Rathgeb:2022yqb}. This scheme, which we just refer to as the $C$-scheme, has the special feature that the QCD $\beta$-function depends only on the universal one- and two-loop coefficients $\beta_0$ and $\beta_1$ and can be written down in closed form. Furthermore, the
QCD hadronization scale $\Lambda_{\rm QCD}$ is identical to the one in the usual $\overline{\rm MS}$ scheme. The simple form of the  $\beta$-function in the $C$-scheme has the important property that the GC renormalon and the associated asymptotic series behavior used in the subtration definition can be written down in a closed and compact form without any truncation concerning subleading asymptotic contributions. This makes the $C$ scheme special and particularly useful for conceptual renormalon studies in full QCD. Our actual numerical studies are, however, still carried out in the usual $\overline{\rm MS}$ strong coupling scheme and are obtained from the $C$-scheme expressions by a finite-order re-expansion. The relevant formulae can be found in the appendix of Ref.~\cite{Benitez-Rathgeb:2022yqb}. For the numerical evaluations of the strong coupling in the complex plane we used the quasi-exact routine from the REvolver library package~\cite{Hoang:2021fhn}. Throughout this work we employ the strong coupling for $n_f=3$ active flavors unless stated otherwise.

The content of this paper is as follows:
In Sec.~\ref{sec:Notation} we set up our notations and briefly review the theoretical ingredients relevant to the phenomenological analysis of $\tau$ hadronic spectral function moments as well as the basics of implementing the RF GC scheme.
Section~\ref{sec:GCnormalization} is dedicated to the determination of the GC renormalon norm $N_g$ using three different methods. Two of the methods are already known from the literature, and the third one is new. The constructive elements of the new method are the improvements in the perturbative behavior for the CIPT and FOPT expansions for GCS and GCE spectral function moments, and we demonstrate that it constitutes a very powerful and reliable method. In Sec.~\ref{sec:tests} we finally put the RF GC to a practical test by applying it in the context of two realistic phenomenological strong coupling determinations following the analyses of Refs.~\cite{Pich:2016bdg,Boito:2020xli}. Here we only rely on the perturbative input that is commonly used in such phenomenological analyses, and we in particular account for the uncertainty in the norm $N_g$ and variations of the IR factorization scale $R$.
In Sec.~\ref{sec:conclusions} we conclude. There are a number of appendices.
In App.~\ref{app:normconventions} we review the different main conventions concerning the definitions used for the renormalon calculus in the literature. The information provided there allows for an easy conversion of our convention for the GC renormalon norm $N_g$ to other conventions used in the literature.
While the main focus of our work lies on the phenomenologically relevant case of $n_f=3$ active quark flavors, our analyses on the determination of the GC renormalon norm $N_g$ can also be applied for the quenched approximation $n_f=0$, for which a number of dedicated analyses are available in the literature. In App.~\ref{sec:GCquenched} we compare our results to these previous determinations. We find a particularly large discrepancy to the results from Ref.~\cite{Bali:2014fea} based on lattice perturbation theory, which we comment on in App.~\ref{sec:commentBalietal}.
\newpage

\section{Notation, Strong Coupling and Renormalon Calculus}
\label{sec:Notation}

\subsection{Theoretical Setup}
\label{sec:theorysetup}

The total hadronic $\tau$ decay rate can be separated into three different components. The contributions from the light-quark ($\bar ud$) current, which can be split into a vector and an axial-vector part, $R_{\tau,V}$ and $R_{\tau,A}$, respectively, and a contribution with net strangeness, $R_{\tau, S}$. In studies aimed at the extraction of the strong coupling, one commonly focuses on the  light-quark contributions, which receive only tiny quark-mass corrections in the theoretical descriptions that can be neglected for all practical purposes. Here we follow this strategy and focus only on the light-quark components.

On the theory side, the description of the $\tau$ hadronic decay  spectral function moments receives contributions from the four two-point $q\bar q$ correlators: vector, axial-vector, scalar, and pseudo-scalar. The contributions of the latter two  are  suppressed by at least two powers of the light-quark masses and are very small.
 Because of chiral symmetry, the perturbative contributions for the vector ($V$) and axial-vector ($A$) currents, which are obtained in the chiral limit, are the same. The nonperturbative contributions, on the other hand, are different for the two currents.

The $V$ and $A$ correlation functions $\Pi^{\mu\nu}_{V/A}(p)$ are
defined as
\begin{equation}
\label{Pimunu}
\Pi^{\mu\nu}_{V/A}(p) \,\equiv\,  i\!\int \! {\rm d}x \, e^{ipx} \,
\langle\Omega|\,T\{ j^\mu_{V/A}(x)\,j^\nu_{V/A}(0)^\dagger\}|\Omega\rangle\,,
\end{equation}
where $|\Omega\rangle$ is the physical QCD vacuum and the currents are
$j^\mu_{V/(A)}(x) = :\!\bar u(x)\gamma^\mu(\gamma_5) d(x)\!:$.
The correlators $\Pi^{\mu\nu}_{V/A}(p)$ admit the usual decomposition into transversal,  $\Pi_{V/A}^{(1)}$, and longitudinal, $\Pi_{V/A}^{(0)}$, components. The longitudinal components in  $V$ and $A$ are related to the scalar and pseudoscalar correlators, respectively. We will often work with the combination
$\Pi_{V/A}(p^2) \equiv \Pi_{V/A}^{(1+0)}(p^2) = \Pi_{V/A}^{(1)}(p^2)+\Pi_{V/A}^{(0)}(p^2)$ and we frequently drop the subscripts $V$/$A$ when suitable. These correlators  are not physical quantities because they are ultra-violet (UV) divergent and require a scale and scheme dependent subtraction. Therefore, one often works with the spectral function,
$\rho(s)$, or the (reduced) Adler function, $D(s)$,  which are UV finite and physical. We use the definitions
\begin{equation}
\label{rhoD}
\rho(s)\equiv \frac{1}{\pi}{\rm Im}\,\Pi(s+i0),\qquad
\frac{1}{4\pi^2}\big[1+D(s)\big] \,\equiv\, -\,s\,
\frac{\rm d}{{\rm d}s}\,\Pi(s) \,.
\end{equation}

In the analysis of hadronic $\tau$ spectral function moments the experimental spectral functions are integrated over weight functions. They can be written, in general, as
\begin{equation}
\label{eq:Rexperimental}
R_{V/A}^{(w)}(s_0) \, =\, 12 \pi^2 \,S_{\rm ew}\,|V_{ud}|^2\,
\int\limits_0^{s_0} \frac{{\rm d}s}{s_0}  w({\textstyle \frac{s}{s_0}}) \left[\,  \rho_{V/A}(s) -\frac{2s/s_0}{1+2s/s_0}\rho_{V/A}^{(0)}(s) \right],
\end{equation}
where the weight function $w(s)$  can be any analytic function; in practice one most often chooses polynomials.  To obtain the physical decay rate, one must set $s_0=m_\tau^2$ and use the weight function determined by the kinematics  of the hadronic $\tau$ decay~\cite{Benitez-Rathgeb:2022yqb,Beneke:2008ad}. Smaller values of the upper limit in the integration, $s_0< m_\tau^2$, are also frequently used in the literature as a means to further constrain the theoretical description~\cite{Boito:2020xli,Boito:2014sta,Boito:2012cr,Maltman:2008nf}, but it is important to keep $s_0$ within the perturbative regime.

The different contributions to the theoretical counterpart of Eq.~(\ref{eq:Rexperimental}) can be decomposed in the following way
\begin{equation}
\label{eq:momdef}
R_{V/A}^{(w)}(s_0) \, =\, \frac{N_c}{2} \,S_{\rm ew}\,|V_{ud}|^2 \Big[\,
\delta^{\rm tree}_{w} + \delta^{(0)}_{w}(s_0)  +
\sum_{d\geq 4}\delta^{(d)}_{w,V/A}(s_0) +\delta_{w,V/A}^{(\rm DV)}(s_0)\Big] \,,
\end{equation}
where $N_c=3$, $S_{\rm ew}=1.0201(3)$ are factorizable electroweak corrections~\cite{Marciano:1988vm,Braaten:1990ef,Erler:2002mv} (which is the value used in Refs.~\cite{Pich:2016bdg,Boito:2020xli}), and $V_{ud}$ is the CKM matrix element. The tree-level contribution is represented by $\delta_{w}^{\rm tree}$. The perturbative QCD corrections, exactly known up to 5 loops, $\mathcal{O}(\alpha_s^4)$, are encoded in $\delta_w^{(0)}$.  The terms $\delta_{w,V/A}^{(d)}$ contain the  nonperturbative corrections arising from the power suppressed contributions of OPE condensates of dimension $d$. Finally, the duality violation (DV) contributions, that  account for nonperturbative corrections that are not captured through the OPE~\cite{Cata:2005zj,Cata:2008ye,Boito:2017cnp} in the region close to the Minkowskian real $s$-axis, are encoded in  $\delta_{w,V/A}^{(\rm DV)}$.

The perturbative contribution, $\delta_w^{(0)}$, was extensively discussed in Part I of this work~\cite{Benitez-Rathgeb:2022yqb}. A review of these results is given in Sec.~\ref{sec:paper1summary}. Here we focus on the nonperturbative terms. The nonperturbative corrections arising from the OPE condensates, using analyticity properties  of the correlators and of the Adler function, can be written  in terms of  contour integrals~\cite{Braaten:1991qm,Beneke:2008ad} ($x=s/s_0$)
\begin{equation}
\label{eq:DOPEmoments}
\sum_{d\geq 4}\delta_{w,V/A}^{(d)}(s_0) =
 4\pi i \ointctrclockwise\limits_{|s|=s_0} \!\!\frac{{\rm d}s}{s_0}  w({\textstyle \frac{s}{s_0}}) \Pi^{\rm OPE}_{V/A}(s)
 = \frac{1}{2\pi i}\,
%\ointctrclockwise\limits_{|s|=s_0} \!\! \frac{{\rm d}s}{s}\,W({\textstyle \frac{s}{s_0}})\, D^{\rm OPE}_{V/A}(s),
\ointctrclockwise\limits_{|x|=1} \!\! \frac{{\rm d}x}{x}\,W(x)\, D^{\rm OPE}_{V/A}(x s_0)\,,
\end{equation}
where the weight function for the Adler function is $W(x)=2\int_x^1 {\rm d}z \,w(z)$.
The  general form of OPE power corrections  for the Adler function  can be written as~\cite{Shifman:1978bx}
\begin{equation}
\label{eq:DOPE}
D^{\rm OPE}_{V/A}(s) \, = \,
\frac{C_{4,0}(\alpha_s(-s))}{s^2} \langle \bar {\cal O}_{4,0} \rangle  +
\sum\limits_{d=6}^\infty \frac{1}{(-s)^{d/2}} \sum_i  C_{d,i}^{V/A}(\alpha_s(-s)) \langle \bar {\cal O}_{d,\gamma_i}\rangle_{V/A}, \,
\end{equation}
where the condensates $ \langle \bar {\cal O}_{d,\gamma_i}\rangle$ are nonperturbative vacuum matrix elements of gauge invariant dimension $d$ operators built from the light-quark and gluon fields, with anomalous dimension $\gamma_i$. The short-distance information is encoded in the Wilson coefficients $C_{d,i}$ which can be computed in perturbation theory as a series  expansion in powers of $\alpha_s(-s)$. In the massless limit, for $d=4$ the only contribution stems from the gluon condensate but starting from $d=6$ contributions from several different condensates need to be considered for each $d$  --- this is accounted for by the sum in $i$.

In practical applications of the OPE  for the analysis of hadronic $\tau$ spectral function moments, due to the lack of knowledge about
terms with $d\geq 6$  and due to the proliferation of operator matrix elements, an  important approximation is
made~\cite{Davier:2008sk,Davier:2013sfa,Maltman:2008nf,Pich:2016bdg,Boito:2012cr,Boito:2014sta,Boito:2020xli}: no $\alpha_s$
corrections in the Wilson coefficients are considered  and the anomalous dimensions of all operators are neglected, thereby removing
the $\alpha_s$-suppressed logarithmic $s$ dependence from the Wilson coefficients of Eq.~(\ref{eq:DOPE}) which reduce to mere tree-level
constants.  This approximation also implies that effectively only a single condensate term contributes at each $d$. For the gluon condensate correction at $d=4$, we use the common renormalization scheme invariant definition
\begin{equation}
\label{eq:GCdefinition}
\langle \bar{\cal O}_{4,0}\rangle \,=\,\frac{2\pi^2}{3} \,\langle\Omega|
\tilde{\beta}(\alpha_s)
\,G^{\mu\nu}G_{\mu\nu}|\Omega \rangle
\,\equiv\, \frac{2\pi^2}{3}\,\langle \bar G^2\rangle\,,
\end{equation}
where $\tilde{\beta}(\alpha_s)\equiv- 2\beta(\alpha_s)/\beta_0\alpha_s=\alpha_s/\pi +\ldots$.
The ${\cal O}(\alpha_s)$ correction in the Wilson coefficient $C_{4,0}(\alpha_s(-s))$ is known~\cite{Chetyrkin:1985kn}.  The latter has sometimes been
included in phenomenological analyses, see e.g. Refs.~\cite{Davier:2013sfa,Boito:2011qt}. However, the logarithmic $s$ dependence
introduced by this correction  leads to rather small numerical effects for gluon condensate
suppressed (GCS) spectral function moments,\footnote{For the GCS spectral function moments the polynomial weight functions $w(x)$
($W(x)$) do not contain a linear term $x$ (quadratic term $x^2$). In all recent
high-precision $\alpha_s$ determinations from spectral function moments GCS spectral function moments were employed due
to their better perturbative behavior and to suppress the rather large impact of the gluon condensate correction which is currently not
known with good precision~\cite{Gubler:2018ctz}.}  so that the approximation of a constant Wilson coefficient is quite good for the
gluon condensate  and has therefore been adopted in most recent spectral function moment analyses. This motivates the common assumption that such an approximation (including the neglect of anomalous dimensions) is also reliable for $d\geq 6$ and that it does not affect the $\alpha_s$ determinations, see e.g.\ Refs.~\cite{Boito:2011qt,Maltman:2008nf},  but no strict proof of this assumption is available. Finally,
 it is also customary to write the OPE in terms of the correlators $\Pi_{V/A}^{(1+0)}$ (instead of the Adler
function). In this simplified form, the Wilson coefficients and the respective effective condensates with dimension $d$ are
collectively represented by a constant $\mathcal{C}_d$ and  one has
\begin{equation}
\Pi_{V/A}^{\rm OPE} \,=\,  \frac{1}{12} \frac{\langle \bar G^2\rangle}{s^{2}} \,+\, \sum_{d=6}^\infty\, \frac{\mathcal{C}_d^{V/A}}{(-s)^{d/2}}\,,
\end{equation}
which gives
\begin{equation}
\label{sec:DOPEnaive}
D_{V/A}^{\rm OPE} \, =\, \frac{2\pi^2}{3}  \frac{\langle \bar G^2\rangle}{s^{2}} \,+ 2\pi^2\, \sum_{d=6}^\infty \, d \frac{\mathcal{C}_d^{V/A}}{(-s)^{d/2}}\,.
\end{equation}
In this form, Cauchy's theorem for  Eq.~(\ref{eq:DOPEmoments})
implies that a monomial term $x^m$ in the weight function $w(x)$ ($W(x)$) picks up solely the OPE correction with dimension $d=2(m+1)$ ($d=2m$).  For GCS moments the  gluon condensate correction term vanishes.

The DV component of the spectral functions, $\rho_{V/A}^{\rm (DV)}(s)$,  cannot be obtained from first principles, but, for sufficiently large $s$, it can be parametrized under generally accepted assumptions about the QCD spectrum (Regge behaviour and large-$N_c$ limit considerations)~\cite{Cata:2005zj,Cata:2008ye,Boito:2017cnp}. Exploiting the analyticity properties of the correlators, it is possible to write the contributions from DV effects as an integral over the respective spectral function contribution. Since the approximate form for the DVs is only valid for sufficiently large $s$, away from the resonance peaks, this contribution is then written as an integral for $s>s_0$  as~\cite{Cata:2008ye}
\begin{equation}
\delta_{w,V/A}^{({\rm DV})} = - 8 \pi^2 \int_{s_0}^\infty \frac{{\rm d}s}{s_0} w({\textstyle \frac{s}{s_0}}) \rho_{V/A}^{\rm (DV)}(s).
\end{equation}
In  this work, when implementing the analysis setup of Ref.~\cite{Boito:2020xli}, the DV contribution to the spectral functions will be parametrized as
\begin{equation}
	\label{eq:DVparametrization}
	\rho_{V/A}^{\rm (DV)}(s) = e^{-\delta_{V/A}-\gamma_{V/A} s}\sin\left(a_{V/A} + b_{V/A}s   \right),
\end{equation}
which introduces four additional parameters per channel: $\delta_{V/A}$, $\gamma_{V/A}$, $a_{V/A}$, and $b_{V/A}$. Following the work of Ref.~\cite{Boito:2017cnp}, corrections to this ansatz arise in the form of logarithms and $1/s$ suppressed terms (see also Eq.~(3.9) of Ref.~\cite{dEnterria:2022hzv}). These corrections have so far not been included in phenomenological analyses.

\subsection{Renormalon-Free GC Scheme for the Adler Function}
\label{sec:paper1summary}

Employing the $C$-scheme for the strong coupling~\cite{Boito:2016pwf} and in the common approach of using a vanishing IR cutoff,
the perturbative series for the Adler function can be written in the form
\begin{eqnarray}
\label{eq:AdlerD}
\hat D(s) & = &
\sum_{\ell=1}^\infty \, \bar c_{\ell} \,\bar a^\ell(-s)\,,
\end{eqnarray}
where ($\beta_0=11-2n_f/3$, $n_f=3$)
\begin{align}
\label{eq:adef}
\bar c_\ell\, \equiv\, \frac{4^\ell\,\bar c_{\ell,1}}{\beta_0^\ell}\,,
\qquad
\bar a(\mu^2)\,\equiv\, \frac{\beta_0\,\bar \alpha_s(\mu^2)}{4\pi}\,.
\end{align}
The perturbative coefficients are known to ${\cal O}(\alpha_s^4)$ (5-loop)~\cite{Baikov:2008jh} and read
\begin{eqnarray}
\label{cn1}
	\bar c_{1,1} &=& 1 \,, \quad
	\bar c_{2,1} \,=\, \sfrac{299}{24} - 9\zeta_3 \,=\, 1.640 \,, \nn \\
	\vbox{\vskip 6mm}
	\bar c_{3,1} &=& \sfrac{262955}{1296} - \sfrac{779}{4}\zeta_3 +
	\sfrac{75}{2} \zeta_5 \,=\, 7.682 \,, \nn \\
	\vbox{\vskip 6mm}
	\bar c_{4,1} &=& \sfrac{357259199}{93312} - \sfrac{1713103}{432}\zeta_3 +
	\sfrac{4185}{8}\zeta_3^2 + \sfrac{34165}{96}\zeta_5 -
	\sfrac{1995}{16}\zeta_7 \,=\, 61.060 \,.
\end{eqnarray}
As is customary in state-of-the-art analyses of hadronic $\tau$ decay spectral function moments, we in addition employ an estimate for the 6-loop coefficient $\bar c_{5,1}$~\cite{Benitez-Rathgeb:2022yqb},
\begin{equation}
\label{eq:c51C}
\bar c_{5,1} \,=\, 345.4774\pm 140\,,
\end{equation}
which covers all the estimates obtained in the recent literature by dedicated analyses~\cite{Boito:2018rwt,Baikov:2008jh,Beneke:2008ad,Caprini:2019kwp,Jamin:2021qxb}. For the coefficients $\bar c_\ell$ this gives $\bar c_1=0.444$,  $\bar c_2=0.324$, $\bar c_3=0.674$, $\bar c_4=2.382$, $\bar c_5=5.991\pm 2.428$. We note that the inclusion of the coefficient $\bar c_{5,1}$ (and its uncertainty estimate) is not at all a minor aspect of the phenomenological analyses, as it leads to a noticeable reduction of the renormalization scale uncertainties. In this work we  adopt the estimate of Eq.~(\ref{eq:c51C}).

The perturbative contributions to the spectral function moments are then determined in analogy to Eq.~(\ref{eq:DOPEmoments}) via the contour integral
\begin{align}
\label{eq:delta0def}
\delta^{(0)}_{w}(s_0)
\, =\,
\frac{1}{2\pi i}\, \ointctrclockwise\limits_{|x|=1}  \!\!\frac{{\rm d}x}{x}\,W(x)\, \hat D(x s_0)\,.
\end{align}
For the CIPT method~\cite{Pivovarov:1991rh} to determine the perturbation series $\delta^{(0)}_{w}(s_0)$ the expansion of the Adler function shown in Eq.~(\ref{eq:AdlerD}) is used, and the contour integration is carried out over powers of the complex-valued strong coupling $\alpha_s(-s)$. The CIPT expansion series arises from truncating the partial sum in Eq.~(\ref{eq:AdlerD}). For the FOPT method the Adler function shown in Eq.~(\ref{eq:AdlerD}) is first reexpanded in powers of $\alpha_s(s_0)$. The contour integration is carried out over the resulting polynomials of $\ln(-s/s_0)$  which multiply the powers of $\alpha_s(s_0)$. The FOPT expansion series arises from truncating the partial sum in powers of $\alpha_s(s_0)$. From the perspective of the FOPT expansion, the CIPT method resums powers of $\ln(e^{\pm i\pi})=\pm i\pi$ terms. For the implementation of renormalization scale variations the CIPT and FOPT methods are actually expansions in powers of $\alpha_s(-\xi s)$ and $\alpha_s(\xi s_0)$, respectively. In our analysis we  in general use the canonical variation interval $1/2\le\xi\le2$.

In the common approach of using a vanishing IR cutoff the coefficients $\bar c_\ell$ contain IR renormalon contributions which cause the Adler function perturbation series to be asymptotic. The different kinds of divergent contributions are compensated by associated divergent order-by-order contributions in the condensates of the OPE corrections shown in Eq.~(\ref{eq:DOPE}). We call this IR factorization scheme for the condensates $\overline{\rm MS}$, which is indicated by the bar over the condensate operators. The explicit form of the GC OPE correction is
\begin{eqnarray}
\label{eq:AdlerOPEGCv2}
	\delta  D^{\rm OPE}_{4,0}(s) & = &
	\frac{1}{s^2} \frac{2\pi^2}{3}\,\Big[ 1 -\frac{22}{81} \, \bar a(-s) + \ldots \,\Big]\, \langle \bar G^2\rangle \,,
\end{eqnarray}
where we have shown the ${\cal O}(\alpha_s)$ corrections to the Wilson coefficient in the renormalization-scale invariant notation of the GC matrix element $\langle \bar G^2\rangle$. The yet unknown higher order corrections are indicated by the ellipses.
The analytic order-dependent form of the IR renormalon contributions contained in the coefficients $\bar c_\ell$ associated to the GC in the $\overline{\rm MS}$ scheme can be unambiguously quantified using the renormalon calculus and can be obtained from the coefficients of the asymptotic series expression
\begin{equation}
\label{eq:AdlerseriesOPEterm}
	\delta  \hat D_{4,0}(s) \, =  \,\frac{2\pi^2}{3}\,N_g\, [ 1 -\frac{22}{81} \, \bar a(-s)
	+ \ldots  ]\,
	\sum_{n=1}^\infty \,
	r_{n}^{(4,0)} \,\bar a(-s)^n\,,
\end{equation}
where
\begin{equation}\label{eq:rn4zero}
	r_{\ell}^{(4,0)}=\Big(\frac{1}{2}\Big)^{\ell+4 \hat b_1}\,\frac{\Gamma(\ell+4 \hat b_1)}{\Gamma(1+4 \hat b_1)}\,,
\end{equation}
and $\hat b_1=\beta_1/2\beta_0^2$, with $\beta_1=102-38\,n_f/3$ being the two-loop QCD $\beta$-function coefficient.
In the $C$-scheme for the strong coupling the form of $r_{\ell}^{(4,0)}$ is exact. The order-dependence proportional to $2^{-\ell}\Gamma(\ell+4 \hat b_1)$  precisely characterizes the divergent asymptotic contributions for the ${\cal O}(\Lambda_{\rm QCD}^4)$ renormalon associated to the GC OPE correction. The only unknown is the exact value of the normalization factor $N_g$, since it cannot be inferred from the renormalon calculus.

The purpose of the RF GC scheme we devised in Part~I~\cite{Benitez-Rathgeb:2022yqb} is to eliminate the contribution of the GC renormalon shown in Eq.~(\ref{eq:AdlerseriesOPEterm}) order-by-order from the perturbative coefficients $\bar c_\ell$ of the Adler function through a scheme change of the GC matrix element while maintaining its scale-invariance and its natural ${\cal O}(\Lambda_{\rm QCD}^4)$ scaling. This is achieved in a first step by the rewriting the $\overline{\rm MS}$ GC $\langle \bar G^2\rangle$ in terms of the renormalon-free GC matrix element $\langle G^2\rangle(R^2)$ which depends on the IR factorization scale $R$:
\begin{equation}
\label{eq:GCIRsubtracted}
\langle \bar G^2\rangle
	\, \equiv \,
	\langle G^2\rangle(R^2)
	\,- \,
	R^4 \, \sum\limits_{\ell=1}^\infty \,
	N_g\,
	r_{\ell}^{(4,0)} \,\bar a^\ell(R^2)\,.
\end{equation}
Here the $\overline{\rm MS}$ GC $\langle \bar G^2\rangle$ has a role analogous to the pole mass in heavy quark physics, while the renormalon-free GC matrix element $\langle G^2\rangle(R^2)$ has the role analogous to a short-distance mass~\cite{Hoang:2020iah,Beneke:2021lkq}. In fact, the definition of $\langle G^2\rangle(R^2)$ is in close analogy to the RS heavy quark mass scheme proposed in Ref.~\cite{Pineda:2001zq}.
Upon insertion in Eq.~(\ref{eq:AdlerOPEGCv2}) the series on the RHS of Eq.~(\ref{eq:GCIRsubtracted}) is combined with the original Adler function series in Eq.~(\ref{eq:AdlerD}). For the IR factorization scale $R$ the proper scaling is $R^2\lesssim m_\tau^2$ for the application to $\tau$ hadronic spectral function moments to ensure that it is still in the perturbative region and to avoid the appearance of large logarithms $\ln(R^2/s_0)$. For the systematic order-by-order cancellation of the divergent renormalon contributions in the Adler function it is essential that the combined series is consistently expanded using the strong coupling at the same renormalization scale either using the CIPT or FOPT renormalization scale prescription described after Eq.~(\ref{eq:delta0def}).

The GC matrix element $\langle G^2\rangle(R^2)$ is scale-dependent and has ${\cal O}(R^4)$ scaling, which renders it not very convenient, since the truncated series values can strongly depend on the values of $R$ and $N_g$. This is remedied in a second step by rewriting $\langle G^2\rangle(R^2)$ in terms of the scale-invariant GC matrix element $\langle G^2\rangle^{\rm RF}$,
\begin{eqnarray}
	\label{eq:GCIRsubtracted2}
	\langle G^2\rangle(R^2)
	& \equiv &
	\langle G^2\rangle^{\rm RF} + N_g \, \bar c_0(R^2)\,,
\end{eqnarray}
where the function $\bar c_0(R^2)$ satisfies the same $R$-evolution equation as the subtraction series on the RHS in Eq.~(\ref{eq:GCIRsubtracted}) and thus of $\langle G^2\rangle(R^2)$. Interestingly, since the divergent subtraction series is associated to a pure ${\cal O}(\Lambda_{\rm QCD}^4)$ renormalon ambiguity, its  $R$-derivative is a convergent series (within its radius of convergence) which can be summed exactly in closed form~\cite{Benitez-Rathgeb:2022yqb},
\begin{eqnarray}
\label{eq:G2Revolution}
\frac{{\rm d}}{{\rm d}\ln R^2}\,\bigg[\,
R^4 \, \sum\limits_{\ell=1}^\infty \,
N_g\,
r_{\ell}^{(4,0)} \,\bar a^\ell(R^2)\,\bigg]
& = &
\frac{{\rm d}}{{\rm d}\ln R^2}\,\langle G^2\rangle(R^2)
\, = \,
\frac{N_g}{2^{4\hat b_1}}\,\frac{R^4\,\bar a(R^2)}{1-2\hat b_1 \bar a(R^2)}
\,.
\end{eqnarray}
A very suitable choice for $\bar c_0(R^2)$ is the Borel sum of the subtraction series itself using the common principle value prescription for the inverse Borel transform integration,
\begin{equation}
\label{eq:subtractclosed}
	\bar c_0(R^2)  \equiv  R^4\,\,\,{\rm PV}\,\int_0^\infty \!
	\frac{ {\rm d} u \,\,e^{-\frac{u}{\bar a(R^2)}}}{(2-u)^{1+4 \hat b_1}}
	 =
	- \frac{R^4\,e^{-\frac{2}{\bar a(R^2)}}}{(\bar a(R^2))^{4\hat b_1}}\,
		{\rm Re}\left[\, e^{4\pi \hat b_1 i}\, \Gamma\Big(\!-4\hat b_1,-\frac{2}{\bar a(R^2)}\,\Big)\,\right]\,,
\end{equation}
where the analytic result of Eq.~(\ref{eq:subtractclosed}) is valid for real-valued $\bar a(R^2)$. Here, $R^4/(2-u)^{1+4 \hat b_1}$ is the Borel (transform) function
of the subtraction series on the RHS of Eq.~(\ref{eq:GCIRsubtracted}) for $N_g=1$ and the integral over the GC renormalon branch cut for $u\ge 2$ is defined using the average of the contour deformation above and below the real axis, referred to as the principal value (PV) prescription.
In total, the expression for the Adler function in the RF GC scheme has the form
\begin{eqnarray}
\label{eq:invBorelDR}
	\hat D^{\rm RF}(s,R^2) & = &
	\frac{1}{s^2}\,\Big[ 1 + \bar c_{4,0}^{(1)} \bar a(-s) \Big]\,\frac{2\pi^2}{3}\,N_g\, \bar c_0(R^2)
	\, + \,
	\sum_{\ell=1}^\infty \, \bar c_{\ell} \,\bar a^\ell(-s)\,  \qquad
	\\
\lefteqn{
	\, - \,
	\Big[ 1 + \bar c_{4,0}^{(1)}\, \bar a(-s)  \Big]\, \frac{2\pi^2}{3}\,N_g\,\frac{R^4}{s^2}
	\sum\limits_{\ell=1}^\infty
	\Big(\frac{1}{2}\Big)^{\ell+4 \hat b_1}\,\frac{\Gamma(\ell+4 \hat b_1)}{\Gamma(1+4 \hat b_1)} \,\bar a^\ell(R^2)\,,
}\nonumber
\end{eqnarray}
where the first term involving $\bar c_0(R^2)$ is treated strictly as a tree-level term (i.e.\ not being reexpanded again and numerically evaluated in the $C$-scheme for the strong coupling).
To obtain the perturbation series, either using the CIPT or the FOPT renormalization scale setting prescription, it is mandatory to consistently expand and truncate the sum of the two series in  $\bar a^\ell(-\xi s)$ and $\bar a^\ell(\xi s_0)$, i.e.\  using the strong coupling at a common renormalization scale. This ensures the systematic removal of the GC renormalon from the Adler function. The GC OPE correction adopts the standard form
\begin{eqnarray}
	\label{eq:AdlerOPEGCBS}
	\delta D^{\rm OPE,RF}_{4,0}(s) & = &
	\frac{1}{s^2}\frac{2\pi^2}{3}\,\Big[ 1 -\frac{22}{81} \, \bar a(-s) \Big]\, \langle G^2\rangle^{\rm RF} \,.
\end{eqnarray}

The RF scheme entails that the difference between the original $\overline{\rm MS}$ GC and our new scale-independent RF  GC, $\langle G^2\rangle^{\rm RF}$, is formally $\mathcal{O}(\bar a_R^{n+1})$. In this sense, the modifications to the GC in  the RF scheme are minimal. Furthermore, the $R$-dependence of $\hat D^{\rm RF}(s,R^2)$
vanishes in the limit of large truncation order and the Borel sum of $\hat D^{\rm RF}(s,R^2)$ (defined with the PV prescription above) agrees with the one of the original $\overline{\rm MS}$ GC scheme Adler function $\hat D(s)$. This means that the RF GC matrix element $\langle G^2\rangle^{\rm RF}$ is defined with respect to the Borel sum in the PV prescription, a scheme that has been considered in the literature before~\cite{Caprini:2009vf,Lee:2010hd,Caprini:2011ya,Bali:2014sja,Ayala:2020pxq,Hayashi:2021vdq,Abbas:2013usa}, but has not been realized in the context of the $\tau$ hadronic spectral function moments in the FOPT and CIPT expansions prior to Ref.~\cite{Benitez-Rathgeb:2022yqb}.

We note that while the dependence on the renormalon norm in Eq.~(\ref{eq:GCIRsubtracted}) is an aspect the RF GC scheme shares with the renormalon-subtraction (RS) approach proposed in Ref.~\cite{Pineda:2001zq}, the additional constructive element of Eq.~(\ref{eq:GCIRsubtracted2}), which renders the RF GC to be scale-invariant, provides the practical advantage that $\langle G^2\rangle^{\rm RF}$ is asymptotically independent of the value of $N_g$.
This means that the value of $\langle G^2\rangle^{\rm RF}$ is more stable with respect to uncertainties in the value of $N_g$.

\section{Determination of the GC Renormalon Norm}
\label{sec:GCnormalization}

In Part I~\cite{Benitez-Rathgeb:2022yqb} we have demonstrated the effectiveness of the RF GC scheme in the context of concrete Borel function models for the all-order perturbation series of the Adler function $\hat D(s)$. For these models the value of the GC renormalon norm $N_g$ was known exactly. We also considered a multi-renormalon model for the Adler function in full QCD, following the construction of Beneke and Jamin~\cite{Beneke:2008ad}, which uses as input the known 5-loop coefficients of Eq.(\ref{cn1}) and the central value for $\bar c_{5,1}$ in Eq.~(\ref{eq:c51C}). The Borel function model was updated to account for the now known 5-loop coefficient of the QCD $\beta$-function and the use of the $C$-scheme, see App.~A of Part~I. The construction principle of the Beneke-Jamin model
is based on the natural assumption that the IR and UV renormalon terms closest to the origin are sufficient to achieve a realistic description of the true Borel function. Because the known perturbative corrections to Adler function are fully consistent with a sizeable GC renormalon norm, the model predicts a sizeable value for $N_g$. We obtained the concrete value $N_g=0.64$ in Part~I. (See App.~\ref{app:normconventions} for how our renormalon norm convention is related to others employed in the recent literature.)

This result and the constructive elements of the Beneke-Jamin model are compatible with the
proposition that the GC renormalon gives a sizeable contribution to the known QCD corrections of the Adler function at the 5-loop level.
In this section we apply this proposition to also quantify the norm $N_g$ with a realistic uncertainty. The uncertainty in $N_g$ will then enter our phenomeological analyses in Sec.~\ref{sec:tests} as an additional parametric error. We carry out our analysis for $n_f=3$ active flavors, which is the flavor number relevant for the analysis of $\tau$ hadronic spectral functions.

In the following three subsections we use three different methods to determine $N_g$. In Sec.~\ref{sec:mr-model} we discuss different modifications of the multi-renormalon model compatible with the naturalness assumption mentioned above to assess the stability of $N_g$ under these variations. In Sec.~\ref{sec:cmapproach} we examine the `conformal mapping approach' introduced by  Lee to determine $N_g$ for $n_f=3$ active flavors in Ref.~\cite{Lee:2011te}. He used the known Taylor expansion terms of the Euclidean Adler function's Borel function $B[\hat D](u)$ at 5 loops and a conformal mapping to relocate the gluon condensate renormalon branch point originally located at $u=2$ into the unit circle such that it becomes the branch point closest to the origin. This allows to determine the norm through a Taylor expansion in the mapped variable around the origin. We show that  the concrete form of Lee's conformal transformation exhibits a rather slow convergence and substantially undershoots the norm at the 5 loop level. We devise improved conformal transformations following
the works of~\cite{Cvetic:2001sn,Contreras:2002kf,Campanario:2005np,Caprini:2009vf}, where the convergence is accelerated and the results at 5 loops are more reliable. Finally, in Sec.~\ref{sec:optimalsubtraction} we discuss a novel method, called `optimal subtraction method', where $N_g$ is determined from a minimization procedure that encodes the two improvements the RF GC scheme achieves for FOPT and CIPT expansion series for GCS and GCE spectral function moments. These improvements are that the CIPT-FOPT discrepancy for GCS moments is removed and that the perturbative convergence for GCE moments is substantially improved~\cite{Benitez-Rathgeb:2022yqb}.
We find that all three methods provide consistent estimates for $N_g$ compatible with $0.64$ and with uncertainties of around $30$-$40\%$.
We adopt the outcome of the optimal subtraction method, which is given in Eq.~(\ref{eq:Ngoptsub}), as our final result for $N_g$ because it does not rely on the estimate of the 6-loop coefficient $\bar c_{5,1}$.

Interestingly, in the quenched approximation the GC renormalon norm $N_g^{(n_f=0)}$ was determined previously in two dedicated analyses by Lee~\cite{Lee:2011te} and Bali~et~al.~\cite{Bali:2014fea}. Since our methods can also be applied in the quenched approximation and because these two analyses quoted incompatible results, we take the opportunity to apply our three methods to also determine $N_g^{(n_f=0)}$.
The results are given and discussed in App~\ref{sec:GCquenched}. In App.~\ref{sec:commentBalietal} we also show that the incomplete approximate knowledge on the description of the large-order asymptotic behavior of the gluon condensate renormalon series in the lattice strong coupling scheme may be the reason, why the result for $N_g^{(n_f=0)}$ obtained by Bali~et~al.\ is so much larger than the one obtained by Lee.

\subsection{Multi-Renormalon Borel Function Model Approach}
\label{sec:mr-model}

Even though the concrete form of the Borel function of the perturbative Adler function $B(u)$ is unknown, the structure of the non-analytic contributions related to IR and UV renormalons is known  due to a one-to-one association to the terms in the OPE (see Eq.~(\ref{eq:DOPE})) for IR renormalons and to higher-dimensional operator insertions related to integrated out short-distance quantum corrections for UV renormalons. To the extent that the QCD $\beta$-function, the anomalous dimensions of the operators and the Wilson coefficients are known, the associated non-analytic contributions can be quantified unambiguously fixing the order-by-order relative behavior of the renormalon contributions in the perturbative coefficients~\cite{Gross:1974jv,tHooft:1977xjm,David:1983gz,Mueller:1984vh,Beneke:1998ui}, see also Sec.~2.3 in Part~I. The Borel function contains a linear combination of such IR and UV renormalon terms plus terms that are fully analytic in the $u$-plane.\footnote{There are other non-analytic terms such as instanton contributions, which have negligible numerical effects and are therefore not relevant for the discussions here.} The non-analytic branch points for IR (UV) renormalons are located on the positive (negative) real $u$-axis and related to equal-sign (sign-alternating) asymptotic series contribution. The analytic term is related to a series contribution with a finite radius of convergence. Non-analytic renormalon terms with branch points further away from the origin are associated to high-dimensional and more power-suppressed contributions.
The information that is a priori unknown is the normalization of the different non-analytic IR or UV contributions.

The previously mentioned proposition and naturalness assumption implies that the renormalon terms closest to the origin in the Borel plane are sufficient to achieve a good approximate description of the Adler function's Borel function and that the known perturbative coefficients of the Adler function are sufficient to approximately determine their normalization. The `multi-renormalon' model construction devised by Beneke and Jamin~\cite{Beneke:2008ad} follows this proposition. The model includes a term related to the gluon condensate OPE correction displayed in Eq.~(\ref{eq:AdlerOPEGCv2}), which has a branch point at $u=2$ and is the IR renormalon located closest to the origin. Furthermore it contains a term associated to the dimension-6 OPE correction according to the simplified expression in Eq.~(\ref{sec:DOPEnaive}) with a branch point at $u=3$ as well as a term accounting for the leading UV renormalon with a branch point at $u=-1$ and an anomalous dimension consistent with the leading double-pole structure known from the large-$\beta_0$ approximation~\cite{Broadhurst:1992si} (see Sec.~4 of part~I). Since the QCD series terms for the Euclidean Adler function up to ${\cal O}(\alpha_s^4)$ do not show any significant sign alternation in either the $\overline{\rm MS}$ or $C$  schemes for the strong coupling, the normalization of the UV renormalons comes out very small, so that the inclusion of the leading UV renormalon term is sufficient~\cite{Beneke:2008ad}.
The model is supplemented by a linear polynomial in $u$ to account for the analytic contribution.

In the analyses of Borel function models of the Adler function $\hat D(s)$~\cite{Beneke:2008ad,Beneke:2012vb,Jamin:2021qxb}
it is common practice to consider the Borel (transform) function,  $B[\hat{D}(s)](u)$, to be defined with respect to the coupling for the renormalization scale $\mu^2=-s$. We adopt this convention here as well and furthermore use the $C$-scheme for the strong coupling. We also refer the reader to App.~\ref{app:normconventions} for the precise definition of the Borel function we employ in this work (and its relation to other conventions used in the literature) and to Ref.~\cite{Boito:2016pwf} for the $C$-scheme.

In this context the Beneke-Jamin model (which was employed in this form in Part~I) adopts the concrete form\footnote{\label{ftn:intbyparts} Through integration-by-parts the ${\cal O}(\bar a^n)$ term of the form $\frac{\bar a^n}{(p-u)^\gamma}$ in the Borel function is equivalent to a term of the form
$\frac{\Gamma(\gamma - n)}{\Gamma(\gamma)(p - u)^{\gamma - n}}-\sum_{i=0}^{n-1}p^{n-\gamma-i}\frac{\Gamma(\gamma-n+i)}{\Gamma(\gamma)\Gamma(1+i)}u^i$,
which is an alternative notation frequently used in the literature. For a more detailed discussion of this point see also Sec.~2.3 of Part I.}
\begin{align}
\label{eq:Bmodel}
\begin{split}
B[\hat{D}(s)]_{\rm mr}(u) &= b^{(0)} + b^{(1)}u + \frac{2\pi^2}{3}\frac{N_{g}
	\left[1-\frac{22}{81}\bar{a}(-s)\right]}{(2-u)^{1+4\hat b_1}} + \frac{N_6}{(3-u)^{1+6\hat b_1}} + \frac{N_{-2}}{(1+u)^{2-2\hat b_1}}\,.
\end{split}
\end{align}
The three renormalon norms $N_{g}$, $N_{6}$ and $N_{-2}$ and the two polynomial coefficients $b^{(0,1)}$ are then determined from the Adler function coefficients up to $\bar c_{5,1}$, which imply that the first 5 terms of the Taylor series for $B[\hat{D}(s)]_{\rm mr}(u)$ read $\sum_{n=1}^5 \bar c_n/\Gamma(n) u^{n-1}$, see Eq.~(\ref{eq:adef}). Using  $\bar c_{5,1} \,=\, 345.477$ and $\hat b_1=32/81=0.395$ we obtain $N_g=0.64$, $N_{6,0}= -15.65$, $N_{-2}= -0.027$, $b^{(0)}= 0.154$ and $b^{(1)}=0.008$, which is the set of values related to the GC norm value mentioned above in the introduction.
It is the idea of the `renormalon model approach' to determine $N_g$ with an uncertainty by considering several modifications in the construction of $B[\hat{D}(s)]_{\rm mr}(u)$ that are consistent with the central proposition and the naturalness assumption.

Let us first consider the impact of the uncertainty on the 6-loop coefficient $\bar c_{5,1}$, see Eq.~(\ref{eq:c51C}).
Changing the coefficient by $\pm 140$ we obtain $N_g=0.64\pm 0.27$ which corresponds to a relative variation of $43\%$. This uncertainty is irreducible in the context of only having estimates for the 6-loop coefficient $\bar c_{5,1}$ using the Borel model approach. It is now interesting examining how the uncertainties related to the structure of the model itself can affect the value of $N_g$.
Let us study the possible impact of the ${\cal O}(\bar a^2)$ correction in the Wilson coefficient of the GC term. Given that the known ${\cal O}(\bar a)$ correction has the coefficient $-22/81\approx 0.27$  and that the  perturbative coefficients $\bar c_\ell$ of the Adler function are very nicely behaved with coefficients well below $0.5$ up to ${\cal O}(\bar a^2)$, see the text below Eq.~(\ref{eq:c51C}), it is reasonable to assume that the size of the yet unknown ${\cal O}(\bar a^2)$ coefficient does not exceed $0.5$ as well. We thus consider a modification of the GC term shown in Eq.~(\ref{eq:Bmodel}) by considering a Wilson coefficient of the form $[1-\frac{22}{81}\bar{a}(-s)+ \delta \,\bar{a}^2(-s)]$ with $-0.5\le\delta\le 0.5$. This leads to the result $N_g=0.64^{+0.11}_{-0.08}$ which corresponds to a relative variation of $+18\%$ and $-13\%$. These variations are much smaller than that coming from $\bar c_{5,1}$. Using the modifications $[1+ \delta \,\bar{a}(-s)]$ for the Wilson coefficient of the $d=6$ ($N_6$) IR renormalon and the UV renormalon ($N_{-2}$) terms with the same $\delta$-variation, we obtain $N_g=0.64^{+0.09}_{-0.03}$ and $N_g=0.64^{+0.01}_{-0.01}$, which are even smaller variations.
Next, let us consider two other structural modifications of the Borel model of Eq.~(\ref{eq:Bmodel}). The fact that the linear coefficient $b^{(1)}$ for the default model of Eq.~(\ref{eq:Bmodel}) is rather small indicates that the contributions coming from the renormalon terms already accounts nicely for all corrections at ${\cal O}(\alpha_s^2)$ and beyond.
We therefore drop the linear term $b^{(1)}u$ and include a $d=8$ renormalon term of the form $N_8/(4-u)^{1+8\hat b_1}$. This yields $N_g=0.68$ which differs only by $6\%$ from the norm obtained from the original model. If we include instead of the $d=8$ renormalon term an additional $u=-2$ UV renormalon term $N_{-4}/(2+u)^{2-4\hat b_1}$ we obtain a result for $N_g$ that only differs by $0.5\%$.

We see that the unknown 6-loop coefficient $\bar c_{5,1}$ represents the largest source of uncertainty of $\pm 40\%$ when determining the GC renormalon norm $N_g$ using the Borel model method.  Increasing the error in $\bar c_{5,1}$ beyond the one adopted in Eq.~(\ref{eq:c51C}) would further increase this uncertainty.\footnote{ In Ref.~\cite{Pich:2016bdg}, see footnote~\ref{ftnt:c51}, a much larger uncertainty is used. This larger uncertainty is much more conservative but is disfavoured by recent dedicated analyses~\cite{Baikov:2008jh,Boito:2018rwt,Caprini:2019kwp,Jamin:2021qxb}, and is also incompatible with our basic proposition that the GC renormalon has a sizeable contribution to the known 5-loop correction, see Ref.~\cite{Beneke:2008ad,Jamin:2021qxb}.} Structural uncertainties related to the form of the Borel, model related to the unkown higher order corrections to the Wilson coefficients or to the form of the renormalons that are accounted for, lead to much smaller effects. Among these subleading sources of uncertainty, the still unknown two-loop correction to the GC Wilson coefficient is the largest with around $15$ to $20\%$. Modifications related to IR renormalon terms associated to condensates beyond dimension-4 or to the UV renormalons (which are known to be small due to the absense of sign alternating contributions) have significantly smaller effects. These observations are consistent with the assumptions entailed in our proposition and the naturalness of the  Adler function's renormalon structure.
It should be noted, however, that the possibilities to change the form of the Borel model are certainly not exhausted with the  modifications we discussed above.  Overall, we therefore consider the outcome of this analysis primarily as a useful starting point and
it is appropriate to consider also the two alternative methods discussed in the following subsections.

\subsection{Conformal Mapping Approach}
\label{sec:cmapproach}

The conformal mapping approach~\cite{Lee:2011te} to determine the GC renormalon norm $N_g$ uses the fact that the function
\begin{align}
\label{eq:Btilde}
\tilde B(u)\equiv \frac{3\,(2-u)^{1+4\hat b_1}}{2\pi^2} \, B[\hat{D}(s)](u)
\end{align}
is analytic in the vicinity around the point $u=2$ as far as gluon condensate renormalon in $B[\hat{D}(s)](u)$ is concerned. It has, however, still a branch point at $u=2$ from
$(2-u)^{1+4\hat b_1}$ multiplying the other renormalons or the analytic contributions. Furthermore, the radius of convergence of the $u^n$ Taylor series of $(2-u)^{1+4\hat b_1}$ is $2$, but the series converges to zero for $u=2$ because $1+4\hat b_1$ is positive.
 The idea of the conformal mapping approach is that one can apply conformal transformations $z=f(u)$ such that the origin is unchanged, i.e.\ $f(0)=0$, and the point $z_2\equiv f(2)$ is closer to the origin than any of the other mapped (and unbounded) singular points $\{f(3), f(4),\ldots, f(-1), f(-2), \dots\}$ related to the other IR and UV renormalons. It is then possible to determine $N_g$ from that Taylor series evaluated at $z=z_2$. The condition $f(0)=0$ ensures that the $n$-th term in this Taylor expansion is determined from terms in the Adler function's perturbation series up to ${\cal O}(\alpha_s^n)$, i.e.\ from $\bar c_{1,1}, \ldots, \bar c_{n,1}$. This means that the Taylor series can be calculated based purely on perturbation theory and that there is no need to consider the reconstruction of the Borel function. The only information needed is the form of the most singular non-analytic structure associated to the GC renormalon. The latter is known exactly from the renormalon calculus and reads $1/(2-u)^{1+4\hat b_1}$. The conformal mapping approach is model-independent in the sense that is does not rely on assumptions about the concrete form of the Borel function. However, it depends on the choice of the conformal transformation and that the resulting series for $N_g$ is already close to saturation for the available orders.

\begin{figure}
	\centering
	\includegraphics[width=\textwidth]{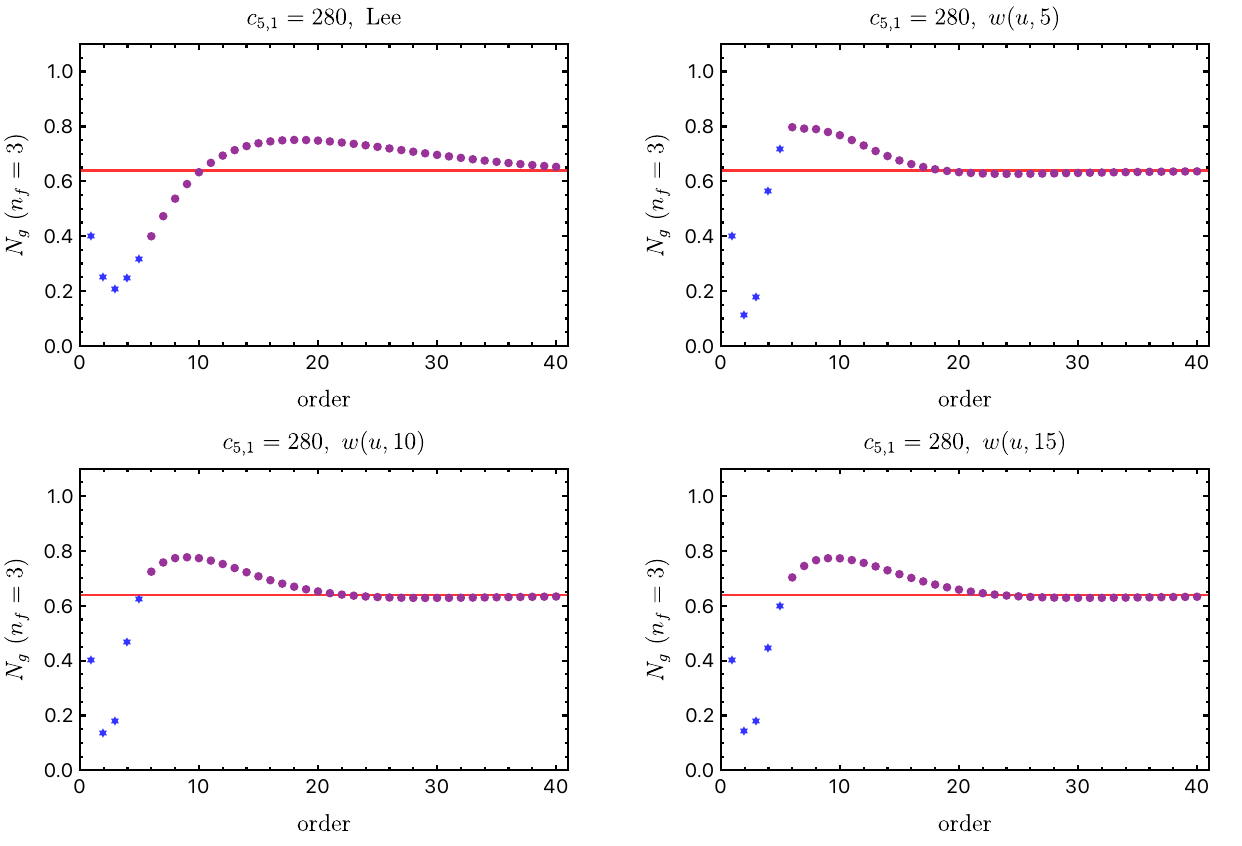}
	\caption{\label{fig:confmapsnf3}
		Series for the $n_f=3$ GC renormalon norm $N_g$ up to order $40$ obtained from the conformal mapping approach using the Borel model of Eq.~(\ref{eq:Bmodel}). The first five orders (indicated by the blue starts) come from the known 5-loop corrections quoted in Eqs.~(\ref{cn1}) and the central estimate for the 6-loop coefficient $\bar c_{5,1}$ of Eq.~(\ref{eq:c51C}). Upper left panel: results based on Lee's mapping of Eq.~(\ref{eq:mappingLee}). The other panels show results for the mapping function $w(u,p)$ of Eq.~(\ref{eq:mappingCaprini}) for $p=5, 10, 15$.}
\end{figure}

In his analysis Lee used the conformal mapping~\cite{Lee:1996yk,Lee:1999ws,Lee:2011te}
\begin{align}
\label{eq:mappingLee}
z=\frac{u}{(1+u)}\,,
\end{align}
which leads to $z_2=2/3\approx 0.667$. Here the singular point second closest to the origin is $r_s=0.75$ and given by the $d=6$ renormalon branch point at $u=3$. This results in the Taylor series $N_g=0.40 - 0.15 - 0.042 + 0.040 + 0.071=0.32$ at ${\cal O}(\alpha_s^5)$ using the central value $\bar c_{5,1}=345$ of Eq.~(\ref{eq:c51C}) for the last term.\footnote{Lee carried out his analysis in the $\msb$ scheme up to ${\cal O}(\alpha_s^4)$ and determined the GC renormalon norm of the Adler function. Accounting for the conventional factor $2\pi^2/3$ in the GC OPE correction shown in Eq.~(\ref{eq:AdlerOPEGCv2}) he obtained ${\cal N}_g=0.048 - 0.018 - 0.008 + 0.001$ in his convention for the GC renormalon norm. Using the estimate of Eq.~(\ref{eq:c51C}), which corresponds to the  $\msb$ coefficient $c_{5,1}=280$~\cite{Benitez-Rathgeb:2021gvw}, the ${\cal O}(\alpha_s^5)$ term reads $+0.005$. In our convention for the GC norm this corresponds to $N_g=0.40 - 0.15 - 0.065 + 0.008 + 0.043=0.24$.} The value of $z_2$ is quite close to $r_s$, and the series does not yet look to be close to a saturation. This suspicion can be substantiated by testing the method using the multi-renormalon model of Eq.~(\ref{eq:Bmodel}) with the set of parameter values displayed in the text below that equation. Since the model can be considered as a reasonable approximation to the full QCD Borel function, we can check how fast the method converges using the conformal mapping of Eq.~(\ref{eq:mappingLee}). The result as a function of order $n$ is shown in the left upper panel of Fig.~\ref{fig:confmapsnf3}. We see that the series is converging to the correct value of $N_g$. However, the series saturates towards $N_g=0.64$ only for orders $n\gtrsim 40$. For orders $n\lesssim 5$ (shown with blue stars in Fig.~\ref{fig:confmapsnf3}) the series significantly undershoots the correct model value. This is far from satisfying and motivates considering an alternative conformal transformation that leads to an improved saturation behavior.

A class of conformal transformations that turns out to be much more suitable is given by
\begin{align}
\label{eq:mappingCaprini}
w(u,p)= \frac{\sqrt{1+u}-\sqrt{1-\frac{u}{p}}}{\sqrt{1+u}+\sqrt{1-\frac{u}{p}}}\,,
\end{align}
where $p$ is a free parameter. This mapping has been employed for specific values of $p$ in this and other contexts in Refs.~\cite{Caprini:1998wg,Cvetic:2001sn,Cvetic:2001ws} and many other works thereafter. We apply it here for the conformal mapping approach to determine $N_g$. For the values $p=(5,10,15)$ we obtain $z_2=(0.38,0.32,0.30)$ and $r_s=(0.52,0.41,0.38)$ for the singular points second closest to the origin, which provides an evaluation of the Taylor series closer to the origin and also has a smaller ratio of $z_2/r_s$. In the upper right and the lower panels of Fig.~\ref{fig:confmapsnf3} we display the convergence of the Taylor series for $N_g$ for $p=5,10,15$ in the context of the Borel model already considered for Lee's mapping. We find a substantially improved convergence behavior. The series saturate towards $N_g=0.64$ already for orders $n\gtrsim 20$. At ${\cal O}(\alpha_s^4)$ and ${\cal O}(\alpha_s^5)$, the series values are already within $30\%$ and $10\%$, respectively, of the correct model value, were the ${\cal O}(\alpha_s^4)$ result is always below the true value. Furthermore, the model values coming from orders beyond ${\cal O}(\alpha_s^5)$ do not deviate from $0.64$ by more than $20\%$.  In view of the uncertainty for $N_g$ from the multi-renormalon model approach this is satisfactory.

\begin{table}
	\begin{center}
		\begin{tabular}{ |p{1.5cm}|p{2.25cm}|p{2.45cm}|p{2.45cm}|   }
			\hline
			& $w(u,5)$ & $w(u,10)$ & $w(u,15)$ \\
			\hline
			${\cal O}(\alpha_s^4)$ &$0.57$& $0.47$& $0.45$\\
			${\cal O}(\alpha_s^5)$ &$0.72 \pm 0.24$& $0.63 \pm 0.17$& $0.60 \pm 0.15$\\
			\hline
		\end{tabular}
		\caption{\label{tab:nf3confmapp} Results for the $n_f=3$ GC renormalon norm $N_g$ at ${\cal O}(\alpha_s^4)$ and ${\cal O}(\alpha_s^5)$ . The uncertainties at order $5$ come from varying $\bar c_{5,1}$ within the uncertainty estimate given in Eq.~(\ref{eq:c51C}).}
	\end{center}
\end{table}

For our analysis we extract the value for $N_g$ at ${\cal O}(\alpha_s^4)$ and ${\cal O}(\alpha_s^5)$, where for the latter we include the uncertainty due to the error in $\bar c_{5,1}$ given in Eq.~(\ref{eq:c51C}). The results are shown in Tab.~\ref{tab:nf3confmapp}. Within the uncertainties the ${\cal O}(\alpha_s^5)$ results are nicely compatible with the ${\cal O}(\alpha_s^4)$ results for all values of $p$. Furthermore at ${\cal O}(\alpha_s^5)$ we obtain the smallest uncertainties for $p=15$. We have checked that the results for the three $p$ values we have shown here are representative for all values of $p>3$ and that, in particular, there is no significant change with respect to the $p=15$ result for even larger values of $p$.
Overall, we find that the conformal mapping method confirms the result for $N_g$ we obtained from the Borel model approach with respect to the central value as well as the uncertainty.

\subsection{Optimal Subtraction Approach}
\label{sec:optimalsubtraction}

The third method to determine the GC renormalon norm $N_g$, which we call `optimal subtraction approach', encodes the two major improvements the RF GC scheme achieves over the previously used $\overline{\rm MS}$ GC scheme for the $\tau$ hadronic spectral function moments: (1) the reduction of the CIPT-FOPT discrepancy for GCS spectral function moments and (2) the improvement of the badly behaved perturbation series for GCE spectral function moments. As we have shown in Part~I, for the proper choice of $N_g$ these two types of improvements are realized simultaneously for any possible choice of GCS or GCE spectral function moments.

It is the idea of the optimal subtraction approach to employ an optimization procedure, based on a $\chi^2$-type minimization, which quantifies the improvements (1) and (2) as a function of $N_g$. To explain the construction of the $\chi^2$ function let us write  FOPT (FO) and CIPT (CI) spectral function moment expansion series for a given weight function $w(x)$ and truncated at ${\cal O}(\alpha_s^m)$ in the form
\begin{align}
\label{eq:delta0series}
\delta^{(0),{\rm FO/CI}}_{w,m}(N_g,s_0; \alpha_s(s_0)) \, = \, \sum_{n=0}^m r_{w,n}^{\rm FO/CI}(N_g,R^2,\xi; s_0,\alpha_s(s_0))\,,
\end{align}
where the index $n$ counts the order in the $\overline{\rm MS}$ strong coupling expansion (either in terms of $\alpha_s(\xi s_0)$ or $\alpha_s(-\xi s)$ prior to the contour integration),\footnote{We remind the reader that in Sec.~\ref{sec:paper1summary} we have formulated the RF GC scheme in the context of the $C$-scheme for the strong coupling, but that all concrete phenomenological analyses are carried out in the common $\overline{\rm MS}$ scheme. The values for $\alpha_s(\xi s_0)$, $\alpha_s(-\xi s)$ and $\bar a(R^2)$ (the latter to be used in the function $\bar c_0$) are obtained from the input $\overline{\rm MS}$ value for $\alpha_s(m_\tau^2)$ using the known $\beta$-function coefficients up to 5 loops.} which we collectively refer to as ${\cal O}(\alpha_s^n)$. The known perturbative coefficients in Eq.~(\ref{cn1}) uniquely quantify $r_{w,n}^{\rm FO/CI}$ for $n=1,2,3,4$, and $r_{w,5}^{\rm FO/CI}$ is determined from the estimate (\ref{eq:c51C}) with an uncertainty.
The coefficients $r_{w,n}^{\rm FO/CI}$ depend on the subtraction scale $R$ and the renormalization scaling parameter $\xi$.
In the formal large order $n$ limit, the dependence of the truncated sum $\delta^{(0),{\rm FO/CI}}_{w,m}$ on $R$ and $\xi$ formally vanishes due to renormalization group invariance, which is the reason why we have suppressed the dependence on $R$ and $\xi$ as arguments in $\delta^{(0),{\rm FO/CI}}_{w,m}$. For any finite order $n$, the truncated sum has, however, a residual dependence on $R$ and $\xi$, as it is common in perturbation theory.  Note that the `tree-level' term $r_{w,0}^{\rm FO/CI}$, which arises in the RF GC scheme and is proportional to $N_g\,\bar c_0(R^2)$, is independent of the renormalization scale parameter $\xi$.

Our $\chi^2$ function, which depends on the truncation order $m$ and is constructed from a set of GCS and GCE spectral function moments, consists of two additive parts
\begin{eqnarray}
\chi^2_{m}(N_g) & = & \chi^2_{m,{\rm GCS}}(N_g) + \chi^2_{m,{\rm GCE}}(N_g)\,.
\end{eqnarray}
To keep the expressions compact we have suppressed all arguments except for $N_g$, but the $\chi^2$ functions also depends on $R$, $\xi$, $s_0$ as well as $\alpha_s(s_0)$.
The first term provides a measure for the CIPT-FOPT discrepancy at order $m$ for a set of GCS spectral function moments:
\begin{align}
\label{eq:chi2GCS}
\chi^2_{m,{\rm GCS}}(N_g) = \sum_{i} \Big(\,\delta^{(0),{\rm CI}}_{w_i,m}(N_g) - \delta^{(0),{\rm FO}}_{w_i,m}(N_g) \Big)^2\,.
\end{align}
We remind the reader that for the GCS moments, the polynomial weight functions $w_i$ do not contain a linear $x$ term and that the GC OPE correction is strongly suppressed. For these spectral function moments, the CIPT as well as FOPT expansions already provide well-behaved perturbation series even in the $\overline{\rm MS}$ GC scheme. A contribution from a GCS moment to $\chi^2_{m,{\rm GCS}}$ is small if the discrepancy between the truncated CIPT and FOPT expansion series is small as well.

The second term provides a measure for the quality of convergence for a set of GCE moments:
\begin{align}
\label{eq:chi2GCE}
\chi^2_{m,{\rm GCE}}(N_g) = \sum_{i} \Big(\,r_{w_i,m}^{\rm FO}(N_g) - r_{w_i,m-1}^{\rm FO}(N_g) \Big)^2\,.
\end{align}
We remind the reader that for the GCE moments, the polynomial weight functions $w_i$ contain a linear $x$ term (with a sizeable coefficient) and that GC OPE correction is sizeable. For these spectral function moments, the CIPT as well as FOPT expansion series in the $\overline{\rm MS}$ GC scheme are quite badly behaved~\cite{Beneke:2012vb}, so the difference between order $m$ and order $m-1$ series terms is sizeable. A contribution from a GCE moment to $\chi^2_{m,{\rm GCE}}$ is small if the series is well behaved. Since the CIPT expansion leads to a significant enhancement of the sign-alternation behavior,\footnote{This property of the CIPT expansion is well-known and has been pointed out in Refs.~\cite{Beneke:2008ad,Beneke:2012vb}. We have also described this behavior in Part~I, see Fig.~3.} which can upset artificially the difference between the size of order $m$ and order $m-1$ series terms, we only account for the FOPT series terms for the construction of $\chi^2_{m,{\rm GCE}}$. The properties of the GCS and GCE moment series in the FOPT and CIPT expansions either in the $\overline{\rm MS}$ or the RF GC scheme mentioned above, which motivate the forms of $\chi^2_{m,{\rm GCS}}$ and $\chi^2_{m,{\rm GCE}}$,
have been discussed in detail in Sec.~5 of Part~I (see in particular Fig.~3).

For the construction of $\chi^2_{m,{\rm GCS}}$ we consider five representative GCS weight functions defined from
\begin{equation}\label{eq:PRSmoments}
 w_n(x)\equiv  w^{(2,n)}(x)= (1-x)^2\sum_{k=0}^n (k+1)x^k = 1 -(n+2)x^{n+1}+(n+1)x^{n+2}\,,
\end{equation}
for $n=1,2,3,4,5$, which were
used in the phenomenological analysis in Sec.~5.3 of Ref.~\cite{Pich:2016bdg}. The first weight function in this sequence, $w_1$, is the well-known kinematic weight function relevant for the inclusive hadronic $\tau$ decay rate. The explicit expressions for the five weight functions are
\begin{eqnarray}
w_1(x) & = &  1-3x^2+2x^3 \,,
\nonumber \\
w_2(x) & = & 1-4x^3+3x^4  \,,
\nonumber \\
w_3(x) & = &  1-5x^4+4x^5  \,,
\nonumber \\
w_4(x) & = &  1-6x^5+5x^6  \,,
\nonumber \\
w_5(x) & = &  1-7x^6+6x^7  \,.
\label{eq:explicitPRSmom}
\end{eqnarray}
They are linearly independent and all doubly pinched.
For the construction of $\chi^2_{m,{\rm GCE}}$ we consider moments obtained from the following representative five GCE weight functions:
\begin{eqnarray}
w_6(x) & = & \frac{3}{2}(1-x)^2 = \frac{3}{2} - 3x +\frac{3x^2}{2} \,,
\nonumber \\
w_7(x) & = & (1-x)^2\left(\frac{13}{12}+\frac{5x}{3}\right) = \frac{13}{12}-\frac{x}{2}-\frac{9x^2}{4}+\frac{5x^3}{3} \,,
\nonumber \\
w_8(x) & = & \frac{1}{2}(1-x)^2\left(5-8x\right)=\frac{5}{2} -9x +\frac{21x^2}{2} - 4x^3 \,,
\nonumber \\
w_9(x) & = & (1-x)^2\left(\frac{3}{2} + x - 3x^2 + x^3\right) = \frac{3}{2} - 2x - \frac{7x^2}{2} + 8x^3 - 5x^4 + x^5  \,,
\nonumber \\
w_{10}(x) & = & (1-x)\left(1 - \frac{x^3}{2} + \frac{3x^4}{4}\right) = 1 - x - \frac{x^3}{2} + \frac{5x^4}{4} - \frac{3x^5}{4} \,.
\end{eqnarray}
They are linearly independent and doubly pinched as well (except for $w_{10}$ which is singly pinched), and we have adopted their form such that the coefficients of the linear $x$ terms have different values that is still of order one, i.e.\ neither too small nor too large.
In any case, all 10 weight functions are in principle also useful for phenomenological analyses with suppressed DV corrections and we find compatible results for other choices of moments. Note that the perturbative spectral function moment series associated to the 10 weight functions (either in the FOPT or the CIPT expansion) all sum to values at ${\cal O}(\alpha_s^4)$ and ${\cal O}(\alpha_s^5)$ very close to the kinematic moment series. This ensures that using the same weight for all moments contributing to the $\chi^2_{m}(N_g)$ does not cause any particular bias.

\begin{figure}
	\centering
	\includegraphics[width=\textwidth]{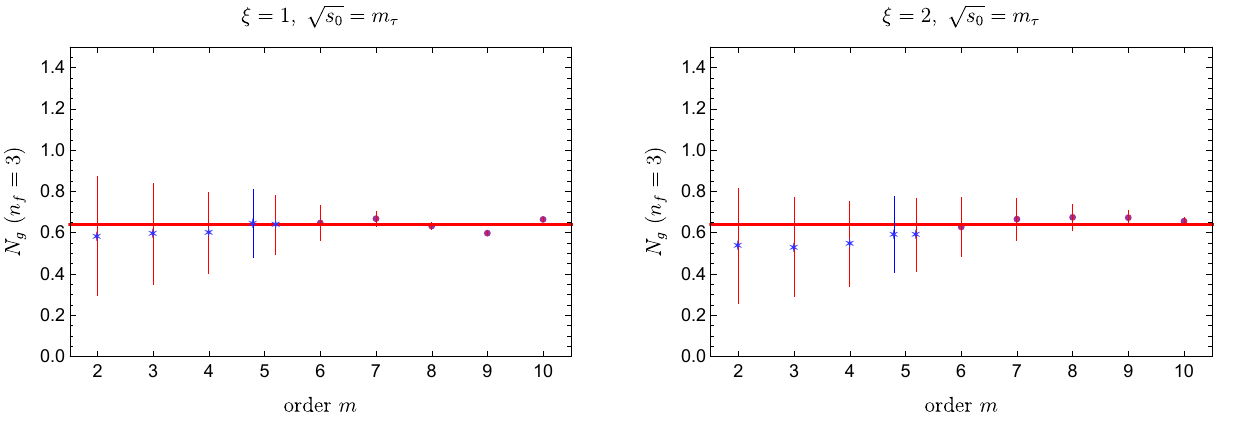}
	\caption{\label{fig:chi2n10}
		Results for the $n_f=3$ GC renormalon norm $N_g$ up to order $m=10$ obtained from the optimal subtraction approach using the Adler function obtained by the Borel model of Eq.~(\ref{eq:Bmodel}). The first five orders, up to ${\cal O}(\alpha_s^5)$, (indicated by the blue starts) come from the known 5-loop corrections quoted in Eqs.~(\ref{cn1}) and the central estimate for the 6-loop coefficient $\bar c_{5,1}$ of Eq.~(\ref{eq:c51C}).
		The red error bars arise from varying the IR factorization scale in the range $0.7\sqrt{s_0}\le R \le \sqrt{s_0}$ and the stars and dots represent the average of the maximal and minimal results at each order in perturbation theory. The blue error bar at order $m=5$ is obtained by conservatively including, in addition to the $R$ variations, independent variations of $\bar{c}_{5,1}$ within the uncertainty estimate of Eq.~(\ref{eq:c51C}). Left panel: Results for $s_0=m_\tau^2$ and $\xi=1$. Right panel: Results for $s_0=m_\tau^2$ and $\xi=2$. }
\end{figure}

To find the best value with an uncertainty for $N_g$ at the truncation order $m$ we determine the minimum of $\chi^2_{m}(N_g)$ for values of the subtraction scale $R$
in the range $0.7\sqrt{s_0}\le R \le \sqrt{s_0}$. For the uncertainty for $N_g$ we adopt half of the range of $N_g$ values that is covered and we take the average of the maximum and minimum values as the central value. As a test of the method, we show the outcome for this analysis for $s_0=m_\tau^2$ up to truncation order $m=10$ in Fig.~\ref{fig:chi2n10} using the series for the Adler function generated by the Borel model of Eq.~(\ref{eq:Bmodel}) for $\xi=1$ (left panel) and $\xi=2$ (right panel). We see that for increasing $m$ the results from this method nicely converge to the correct value for $N_g$ (for the  Borel model this value is $0.64$ and is indicated by the horizontal red line). We also see that for orders $m>6$ the method is slightly more stable for $\xi=2$. This is related to the fact that for strong coupling renormalization scales larger than $\sqrt{s_0}$ the sign-alternating effects the UV renormalons, which have branch points at $u=-1$, are suppressed with respect to the GC renormalon with the branch point at $u=2$. This sign-alternation disturbs the constructive structure of the $\chi^2$ function which exclusively focuses on the impact of the GC renormalon subtraction. Since, eventually, the UV renormalon will dominate the Adler function's perturbation series at very high orders (see Sec.~5 and Fig.~3 of Part~I where the moments for $w_1$ and $w_6$ have been analyzed in great detail), the implementation of the optimal subtraction method we have adopted here cannot be applied to all orders $m$. However, for truncation orders accessible by available or foreseeable calculations the method is perfectly adequate.

For the Borel model in Eq.~(\ref{eq:Bmodel}) the central value for the ${\cal O}(\alpha_s^5)$ coefficient
Eq.~(\ref{eq:c51C}), $\bar c_{5,1}=345$  has been adopted. Interestingly, the additional error on $N_g$ related to the uncertainty on $\bar c_{5,1}$ turns out to be quite small.
To demonstrate that we have at order $m=5$ also determined the range of $N_g$ values by carrying out the $R$ variation described above and {\it in addition} also varied independently $\bar c_{5,1}$ by $\pm 140$. The outcome is shown as the blue error bar in both panels of Fig.~\ref{fig:chi2n10}. We have checked that the outcome of the analysis remains essentially unchanged, if other moments are used that satisfy the same criteria as $w_{1-10}$ or if higher values for $s_0$ are adopted. It should be noted, however, that it is not easy to construct completely different inequivalent sets of analogous moments which are linearly independent when imposing a limit on the maximal power of $x$.

\begin{table}
	\begin{center}
		\begin{tabular}{ |p{2.5cm}|p{2.5cm}|p{2.5cm}|   }
			\hline
			&$\sqrt{s_0}=m_\tau$&$\sqrt{s_0}=3 ~ {\rm GeV}$\\
			\hline
			$m=4  ~ (\xi=1)$& $0.60 \pm 0.20$& $0.51 \pm 0.17$\\
			$m=4  ~ (\xi=2)$& $0.54 \pm 0.21$& $0.55 \pm 0.11$\\
			$m=5  ~ (\xi=1)$& $0.64 \pm 0.16$& $0.50 \pm 0.19$\\
			$m=5  ~ (\xi=2)$& $0.59 \pm 0.18$& $0.52 \pm 0.15$\\
			\hline
		\end{tabular}
		\caption{\label{tab:nf3chi2}
		Results for the $n_f=3$ GC renormalon norm $N_g$ at orders $m=4$ and $5$ obtained from the optimal subtraction approach for $s_0=m_\tau^2, (3\,\mbox{GeV})$, and $\xi=1,2$.
		The uncertainties at order $m=4$ arise from varying the IR factorization scale in the range $0.7\sqrt{s_0}\le R \le \sqrt{s_0}$ and at order $m=5$ from the additional independent variation of $\bar{c}_{5,1}$ within the uncertainty estimate of Eq.~(\ref{eq:c51C}). The central values are the average of the  maximal and minimal results at each order in perturbation theory.}
	\end{center}
\end{table}

In Tab.~\ref{tab:nf3chi2} we display the results of the optimal subtraction method at truncation order $m=4$ and $m=5$ for $\sqrt{s_0}=m_\tau$ and $\sqrt{s_0}=3$~GeV
and using $\xi=1$ and $\xi=2$. For $m=5$ the error on $\bar c_{5,1}$ is included as described in the previous paragraph. All results are equivalent and consistent and we adopt the envelope of the $m=4$ results for $\xi=1$ and $\xi=2$ as our final result:
\begin{equation}
\label{eq:Ngoptsub}
	N_g=0.57\pm 0.23\,.
\end{equation}
The result, which has a relative uncertainty of $40\%$, is fully consistent with the estimates we have obtained from the Borel model and the conformal mapping approaches discussed in Sec.~\ref{sec:mr-model} and \ref{sec:cmapproach} and is furthermore independent of the estimate for the 6-loop coefficient $\bar c_{5,1}$.

The equivalence of the results obtained from the three approaches we have discussed and which all rely on different criteria underlines that the known perturbative coefficients of the Adler function are fully compatible with a sizeable GC renormalon norm in the ranges obtained by our estimates. Within the natural proposition that the GC renormalon has a sizeable contribution to the coefficient at ${\cal O}(\alpha_s^4)$, i.e.\ that the size of $\bar c_{4,1}$ is not just accidentally mimicking this property, our results provide a realistic estimate on the GC renormalon norm $N_g$.
We therefore adopt the result given in Eq.~(\ref{eq:Ngoptsub}) as our final result for the GC renormalon norm.

\section{Impact on Strong Coupling Determinations}
\label{sec:tests}
In order to demonstrate the improvements that can be achieved concerning the CIPT-FOPT discrepancy
when using the RF scheme for the GC in realistic $\alpha_s$   analyses of $\tau$ hadronic spectral function moments, we exemplarily carry out in this section two determinations of $\alpha_s(m_\tau^2)$ following two analysis setups employed in the recent references by Pich and Rodr\'iguez-Sanchez~\cite{Pich:2016bdg} and by Boito, Golterman, Maltman, Peris, Rodrigues,  and Schaaf~\cite{Boito:2020xli}.
These two references are representatives of the two major approaches concerning the treatment of nonperturbative corrections currently used in the literature,  one employing  a truncated version of the traditional parametrization in terms of the OPE corrections, see Refs.~\cite{LeDiberder:1992zhd,Davier:2005xq,Davier:2008sk,Davier:2013sfa}, and one that includes, besides the OPE corrections, DV contributions, see Refs.~\cite{Boito:2011qt,Boito:2012cr,Boito:2014sta}. The two approaches are discussed controversially, see e.g.~Refs.~\cite{Boito:2016oam,Boito:2018yml,Boito:2019iwh} and \cite{Pich:2016bdg,Pich:2022tca}.
For each approach we first employ the CIPT and FOPT expansions in the common $\overline{\text{MS}}$ GC scheme, reproducing the respective published results, and then carry out the analyses in the RF GC scheme, using our result for the GC renormalon norm $N_g$ quoted in Eq.~(\ref{eq:Ngoptsub}). For the analysis in the RF GC scheme, additional uncertainties due to the approximate knowledge of $N_g$ and due to variations of the IR subtraction scale $R$ are accounted for. We find that their numerical impact is relatively small.  This is particularly striking given the quite sizeable uncertainty of $N_g$. This outcome is highly encouraging and underlines the practical value of the RF GC scheme.

We emphasize that the main purpose of the analyses in this section is to demonstrate that the CIPT-FOPT discrepancy, that is present in the $\overline{\text{MS}}$ GC scheme, is significantly reduced in the RF GC scheme, regardless of which approach is employed concerning the treatment of the nonperturbative corrections. In our analyses we use the central value for the estimate of the 6-loop coefficient $\bar c_{5,1}$ shown in Eq.~(\ref{eq:c51C}), for which there is a common agreement in the literature~\cite{Pich:2016bdg,Boito:2020xli,Ayala:2021yct,dEnterria:2022hzv}.\footnote{\label{ftnt:c51}In Ref.~\cite{Pich:2016bdg} $c_{5,1}=275\pm 400$ is used for the 6-loop coefficient using the $\overline{\rm MS}$ scheme for the strong coupling, which corresponds to $\bar c_{5,1}=340.477\pm 400$ in the $C$-scheme. In Ref.~\cite{Boito:2020xli} $c_{5,1}=283\pm 142$ is used, which corresponds to $\bar c_{5,1}=348.477\pm 142$. In Ref.~\cite{Ayala:2021yct} $c_{5,1}=275\pm 63$ is used, which corresponds to $\bar c_{5,1}=340.477\pm 63$.} When assessing theoretical uncertainties arising from  the estimate of $c_{5,1}$, we use the uncertainty quoted in Eq.~(\ref{eq:c51C}). There is no common agreement  for  this uncertainty in the literature  in the context of $\alpha_s$ determinations, but for the purpose of our analysis this aspect is not essential since using the central value is sufficient to demonstrate the improvement.  We adopt the estimate given in Eq.~(\ref{eq:c51C}) as it is consistent with the basic proposition our work is based on and furthermore not overly agressive, see Refs.~\cite{Beneke:2008ad,Jamin:2021qxb}.
Our results show that a higher precision  can in general be reached in $\alpha_s$ determinations from $\tau$ hadronic spectral function moments based on the CIPT and FOPT expansions when the RF GC scheme is employed. The results can be consistently combined in the RF GC scheme, while a combination in the $\overline{\rm MS}$ GC scheme may lead to inconsistent results due to the asymptotic separation. We stress that the analyses carried out in this section should not be interpreted as a supersedure of the results for the strong coupling given in Refs.~\cite{Pich:2016bdg} and \cite{Boito:2020xli} since they carried out many more additional examinations  we do not repeat here due to lack of space.
Furthermore, our results should not be interpreted as an approval (or disapproval) of the methods employed in these references.
A dedicated new strong coupling analysis using $\tau$ hadronic spectral function moments will be carried out by us in a subsequent work.

As far as experimental data is concerned, we note that the LEP experiments, ALEPH~\cite{ALEPH:1998rgl,ALEPH:2005qgp,Davier:2008sk,Davier:2013sfa} and OPAL~\cite{OPAL:1998rrm,Boito:2012cr}, have produced spectral function data from  the analyses of hadronic tau decays for the $V$, $A$, $V+A$, and $V-A$ channels. More recently, with the wealth of data produced by experiments measuring $e^+e^- \to {\rm hadrons}$  in the past decade (mainly aiming at the dispersive calculation of the muon $g-2$), it has become possible to produce a new isovector $V$ spectral function exclusively based on experimental data~\cite{Boito:2020xli}. This has been achieved by extracting subleading channels from recently measured cross-sections for $e^+e^-\to {\rm hadrons}$ using the conserved vector current (CVC) hypothesis. (We remind the reader that the ALEPH and OPAL analyses had to rely on  Monte Carlo simulated data for these subleading channels.) Realistic phenomenological analyses are either based on the 2013 update of the ALEPH spectral functions~\cite{Davier:2013sfa,Pich:2016bdg,Boito:2014sta,Ayala:2021mwc,Ayala:2021yct}, which allow for the analysis of the ``more inclusive'' $V+A$ channel, or on the new purely data based  $V$ spectral function, which has significantly smaller uncertainties near the end-point of the spectrum~\cite{Boito:2020xli}.
In order to follow the works of Refs.~\cite{Pich:2016bdg} and~\cite{Boito:2020xli} we employ,  in Sec.~\ref{sec:truncatedOPE}, the ALEPH $V+A$ data, while in Sec.~\ref{sec:singlemoment} we use the new $V$ spectral function of  Ref.~\cite{Boito:2020xli}.  This introduces an additional difference, besides the treatment of nonperturbative corrections, between the results obtained in these two sections, which makes a consistent combination of them non-trivial.

\subsection[Truncated OPE Analysis:  Multiple Moments at fixed  \texorpdfstring{$s_0$}{s0}]{Truncated OPE Analysis:  Multiple Moments at fixed \boldmath \texorpdfstring{$s_0$}{s0}}
\label{sec:truncatedOPE}

\begin{table}
\begin{center}
\begin{tabular}{|p{1.5cm}|p{1.5cm}|p{1.5cm}|p{1.5cm}|p{1.5cm}|p{1.5cm}|p{1.5cm}|p{1.3cm}|}
\hline
& central value&$\sigma_{\text{exp}}$&$\sigma_{\mu}$&$\sigma_{R}$&$\sigma_{N_g}$&$\sigma_5$&$\sigma_{\text{tot}}$\\
 \hline
 \multicolumn{8}{|c|}{CIPT, $\chi^2$/dof = 0.82, $p$-value = 0.37} \\
 \hline
$\alpha_s(m_{\tau})$&0.3366&0.0042&0.0051&$-$&$-$&0.0024&0.0070\\
$\mathcal{C}_ 6\times 10^{3}$&0.95&0.38&0.44&$-$&$-$&0.012&0.58\\
$\mathcal{C}_ 8\times 10^{3}$&$-$1.07&0.46&0.22&$-$&$-$&0.055&0.51\\
$\mathcal{C}_ {10}\times 10^{3}$&0.22&0.28&0.49&$-$&$-$&0.0052&0.56\\
\hline
 \multicolumn{8}{|c|}{FOPT, $\chi^2$/dof = 1.23, $p$-value = 0.27} \\
 \hline
$\alpha_s(m_{\tau})$&0.3169&0.0031&0.0056&$-$&$-$&0.0014&0.0065\\
$\mathcal{C}_ 6\times 10^{3}$&1.37&0.43&1.22&$-$&$-$&0.076&1.29\\
$\mathcal{C}_ 8\times 10^{3}$&$-$0.95&0.45&1.02&$-$&$-$&0.069&1.12\\
$\mathcal{C}_ {10}\times 10^{3}$&0.43&0.30&0.58&$-$&$-$&0.041&0.66\\
 \hline
 \multicolumn{8}{|c|}{RF GC scheme, CIPT, $\chi^2$/dof = 2.14, $p$-value = 0.14} \\
\hline
$\alpha_s(m_{\tau})$&0.3197&0.0035&0.0047&0.0054&0.0058&0.0017&0.010\\
$\mathcal{C}_ 6\times 10^{3}$&4.96&0.63&0.40&0.70&1.13&0.14&1.53\\
$\mathcal{C}_ 8\times 10^{3}$&$-$3.40&0.54&0.42&0.37&0.76&0.017&1.09\\
$\mathcal{C}_ {10}\times 10^{3}$&1.50&0.33&0.22&0.33&0.41&0.020&0.65\\
\hline
 \multicolumn{8}{|c|}{RF GC scheme, FOPT, $\chi^2$/dof = 1.21, $p$-value = 0.27} \\
\hline
$\alpha_s(m_{\tau})$&0.3169&0.0031&0.0059&0&0&0.0014&0.0068\\
$\mathcal{C}_ 6\times 10^{3}$&1.39&0.43&0.96&0.11&0.0048&0.075&1.06\\
$\mathcal{C}_ 8\times 10^{3}$&$-$0.80&0.45&0.99&0.14&0.059&0.065&1.10\\
$\mathcal{C}_ {10}\times 10^{3}$&0.40&0.30&0.57&0.039&0.014&0.040&0.65\\
\hline
\end{tabular}
\caption{Fitted parameters from the  ALEPH $V+A$ spectral function used in Ref.~\cite{Pich:2016bdg} for CIPT and FOPT in the $\overline{\text{MS}}$ GC scheme and in the RF GC scheme. For the latter we use $N_g=0.57\pm0.23$.  The uncertainty $\sigma_5$ reflects the variation of the 6-loop  coefficient as in Eq.~(\ref{eq:c51C}), while $\sigma_\mu$ and $\sigma_R$ refer to variations of the strong-coupling renormalization scale $\mu$ and  of the  IR subtraction scale $R$ as described in the text. The values for  $\sigma_\mu$, $\sigma_{5}$, $\sigma_R$, and
$\sigma_{N_g}$ are half the difference between the maximum and minimum values obtained from the $\xi$,
$c_{5,1}$, $R$, and $N_g$ variations, respectively, using the central values for the remaining parameters.  The two zeros in the lowest section refer to numbers smaller than $10^{-5}$. The $\mathcal{C}_k$ are given in units of GeV$^k$ and are  denoted   $\mathcal{O}_k$ in Ref.~\cite{Pich:2016bdg}.}
\label{tab:PRSResults}
\end{center}
\end{table}

We start by revisiting  the analysis conducted by Pich~et al.\ in Sec.~5.3 of Ref.~\cite{Pich:2016bdg}. This analysis follows the so-called "truncated OPE strategy" (tOPE) (which is also the basis of the earlier analysis in Ref.~\cite{Davier:2013sfa}) and consists of the following. One chooses a number of linearly independent integrated spectral function moments calculated at $s_0=m_\tau^2$, all based on (at least) doubly-pinched weight functions. The pinching is argued to be sufficient to allow for a neglect of any contribution from  DV corrections. Furthermore, to allow for a fit  procedure, the number of moments must, of course, be larger than the number of parameters in the theory description. Since the pinching and the requirement that the moments are linearly independent bring in additional powers of $x$ in the weight functions $w(x)$ and the associated OPE corrections within the approximation of Eq.~(\ref{sec:DOPEnaive}), it becomes necessary to truncate the tower of OPE condensates in order to have a manageable number of parameters in the fit procedure.  This is based on the argument that the contributions from the neglected higher power OPE corrections   are hierarchical and small.

In the  analysis of Sec.~5.3 of Ref.~\cite{Pich:2016bdg}, the five doubly-pinched weight functions $w_{1,\ldots,5}(x)$ of Eqs.~(\ref{eq:explicitPRSmom}) are used.
(We remind the reader that $w_1(x)$ corresponds to the kinematic moment relevant for the inclusive hadronic $\tau$ decay rate.) Due to the absence of any linear term $x$, all moments are GCS, i.e.\ the GC OPE correction is highly suppressed (and absent in the approximation of Eq.~(\ref{sec:DOPEnaive})), but involve the condensates $\mathcal{C}_{2k}$ for $k=3,\dots,8$.

The four parameters of the fit are $\alpha_s$ as well as the  first three OPE condensates that contribute at leading-
order in the strong  coupling in the approximation of Eq.~(\ref{sec:DOPEnaive}): $\mathcal{C}_6$, $\mathcal{C}_8$, and
$\mathcal{C}_{10}$.  The contributions from condensates of dimension 12, 14, and 16 are neglected, so that for the
moments with $w_{4}(x)$ and $w_{5}(x)$ only the perturbative corrections enter the fit.  In the analysis of Ref.~\cite{Pich:2016bdg},
the  ALEPH $V$, $A$ and $V+A$ data from the update of Ref.~\cite{Davier:2013sfa} were used for three separate fits,
which lead to perfectly equivalent determinations of $\alpha_s(m_\tau^2)$. In the analysis below we therefore only use
the $V+A$ data from that reference. As far as the improvement of the CIPT-FOPT discrepancy in the RF GC scheme is
concerned, the results based on the $V$ or the $A$ data are equivalent.

The results of the fits based on the CIPT and FOPT expansions in the usual $\overline{\text{MS}}$ scheme  for the GC are shown in the upper half of
Tab.~\ref{tab:PRSResults}.
For the strong coupling the results are\footnote{In Ref.~\cite{Pich:2016bdg} additional OPE truncation uncertainties of $0.013$ and $0.012$ for $\alpha_s(m_\tau^2)^{\rm CIPT}$ and $\alpha_s(m_\tau^2)^{\rm FOPT}$, respectively, were added (see their table 8). We do not include these uncertainties in our analysis as our main focus lies on the reconciliation of the CIPT-FOPT discrepancy. }
\begin{equation}
\label{eq:alphastOPEMSbar}
\begin{array}{l}
\alpha_s(m_\tau^2)^{\rm CIPT} \, = \, 0.3366 \,\pm\, 0.0070 \\
\alpha_s(m_\tau^2)^{\rm FOPT} \, = \, 0.3169 \,\pm\, 0.0065
\end{array}\,\qquad \mbox{(tOPE, $\overline{\rm MS}$ GC, ALEPH $V+A$).}
%\qquad\longrightarrow\qquad
%\alpha_s(m_\tau^2) \, = \, 0.3268 \,\pm\, 0.0118
\end{equation}
The central values are obtained using $\bar c_{5,1}=345$ for the 6-loop coefficient.
Apart from the  experimental error from the fits, $\sigma_{\rm  exp}$ (which we treat statistically), we also quote the
theory errors related to the truncation of perturbation theory: $\sigma_\mu$, arising from renormalization scale
variations, and $\sigma_{5}$, from the uncertainty in the estimate of the 6-loop coefficient $\bar c_{5,1}= 345\pm 140$  in Eq.~(\ref{eq:c51C}). For the variation of the renormalization scale
in CIPT, we expand in powers of $\alpha_s(-\xi s)\equiv \alpha_s(-\xi m_\tau^2 \,x)$ prior to carrying out the contour integration and, in FOPT, in powers of $\alpha_s(\xi m_\tau^2)$, with $1/2\leq \xi \leq 2$ and $\xi=1$ being the choice for the central value.
The total uncertainty is obtained by quadratially adding $\sigma_{\rm  exp}$,  $\sigma_\mu$, and $\sigma_5$.
Our results reproduce very well those quoted in the last two lines of table 7 in Ref.~\cite{Pich:2016bdg}.

We see that the results for $\alpha_s(m_\tau^2)$ obtained from the CIPT and FOPT expansion show a difference of about $0.0197$,  which is the typical size of the CIPT-FOPT discrepancy that has been found in Ref.~\cite{Pich:2016bdg} and in many other previous phenomenological analyses based on variants of the tOPE strategy. This is 5 to 6 times larger than the individual experimental uncertainties, and  about 3 times larger than the total uncertainties we obtain for the CIPT and FOPT results for $\alpha_s(m_\tau^2)$ in our analysis. Interestingly, the fit results for the condensates obtained in the CIPT and FOPT analyses are perfectly compatible within uncertainties indicating that the asymptotic separation mostly affects the value of the strong coupling.

In the lower half of Tab.~\ref{tab:PRSResults} we show the corresponding CIPT and FOPT results obtained in the RF GC scheme using  Eq.~(\ref{eq:Ngoptsub}) for the gluon condensate renormalon norm $N_g$ and  $0.7m_\tau\leq R \leq m_\tau$ for the IR subtraction scale variation.  For the strong coupling the results are
\begin{equation}
\label{eq:alphastOPERF}
\begin{array}{l}
\alpha_s(m_\tau^2)^{\rm CIPT} \, = \, 0.3197 \,\pm\, 0.0100 \\
\alpha_s(m_\tau^2)^{\rm FOPT} \, = \, 0.3169 \,\pm\, 0.0068
\end{array} \qquad\mbox{(tOPE, RF GC, ALEPH $V+A$)}.
\end{equation}
The central values are obtained from using $N_g=0.57$ and $R=0.8\, m_\tau$ as well as $\bar c_{5,1}=345$.
Apart from the uncertainties related to the truncation of perturbation theory, $\sigma_\mu$ and  $\sigma_5$, which we have estimated as in the  $\overline{\text{MS}}$ GC scheme, we also quote the uncertainties $\sigma_R$ and $\sigma_{N_g}$ coming from the $R$-variation and the error in $N_g$, respectively.
We see that the
FOPT results are almost unaltered by the subtraction of the GC renormalon and the switch to the RF GC scheme.
 At the same time, the uncertainties $\sigma_R$ and $\sigma_{N_g}$ are extremely small for $\alpha_s$,  and lead to only minor changes of the fit results for the OPE condensates. Overall, switching from the $\overline{\text{MS}}$ to the RF GC scheme for GCS spectral function moments has very little impact for the FOPT expansion. So for the FOPT expansion the suppression of the GC OPE corrections is associated to a supression of the effects induced by switching to the GC scheme. This corroborates the conclusions of Refs.~\cite{Hoang:2020mkw,Hoang:2021nlz} and our observations in Part~I, that the FOPT expansion is consistent with the standard form of the OPE.

The CIPT results, on the other hand, change significantly,
and now the difference of the CIPT and FOPT central values for $\alpha_s(m_\tau^2)$ is reduced to merely $0.0028$, which is smaller than the individual  experimental uncertainties.  In CIPT, the RF GC scheme introduces a new source of theory uncertainty for the $\alpha_s$ value, which now receives non-negligible contributions from  $\sigma_{N_g}$ and $\sigma_R$  even though the numerical size of the GC OPE correction  itself is still strongly suppressed (and even vanishing in the approximation of Eq.~(\ref{sec:DOPEnaive})). This (apparently) paradoxical behavior corroborates the conclusions of Refs.~\cite{Hoang:2020mkw,Hoang:2021nlz} and our observations in Part~I, and illustrates the incompatibility of the CIPT expansion with the standard form of the OPE  as given in Eq.~(\ref{eq:DOPE}) and (\ref{sec:DOPEnaive}), and the need to switch to the RF GC scheme to eliminate the dominating inconsistency that is related to the GC renormalon. It is remarkable, however, that the final uncertainty for $\alpha_s(m_\tau^2)$, which has to account for $\sigma_{N_g}$ and $\sigma_R$, still remains quite small and only grows from $0.0070$ to about $0.010$. This growth in total uncertainty for the CIPT analysis is more than compensated by the gain  in consistency between the FOPT and CIPT analyses. Finally, we observe a change in the central values  of the OPE condensates in CIPT when the RF GC is used. The new results are still marginally compatible with the ones obtained in the  $\overline{\text{MS}}$ GC scheme. We also observe that in the RF GC scheme, the fit quality of the CIPT analysis slightly worsens. Still, the fit is perfectly acceptable with a $p$-value of $14\%$.

Given that the asymptotic separation inherent to the CIPT expansion method in the original $\overline{\rm MS}$ GC scheme is removed in the RF GC scheme, it now makes good sense to determine a combined $\alpha_s$ result, in contrast to the results obtained in the  $\overline{\rm MS}$ GC scheme. We average the FOPT and CIPT results in the RF GC scheme following a prescription given in Ref.~\cite{Pich:2016bdg}: we  take the average of the two individual central values as the final central value and the quadratic sum of the smaller individual uncertainty and half of the central difference as the final combined uncertainty. (We do not reduce the individual errors in the averaging due to the potential correlations in central values.) This results in\footnote{In Ref.~\cite{Pich:2016bdg}, uncertainties of $\pm 400$ were employed for $\bar c_{5,1}$. If we had adopted these variations, $\sigma_5$ would increase to $0.0069$ for CIPT and $0.0039$ for FOPT in the $\overline{\rm MS}$ GC scheme, which would yield
$\alpha_s(m_\tau^2)^{\rm CIPT} =  0.3366 \,\pm\, 0.0096$ and
$\alpha_s(m_\tau^2)^{\rm FOPT} =  0.3169 \,\pm\, 0.0075$. %If we had adopted uncertainties of $\pm 400$ for $\bar c_{5,1}$, $\sigma_5$ would increase to $0.0049$ and CIPT and $0.0039$ for FOPT, which would yield
In the RF GC scheme, a similar increase in uncertainty is observed and the final values would be $\alpha_s(m_\tau^2)^{\rm CIPT} =  0.320 \,\pm\, 0.011$ and  $\alpha_s(m_\tau^2)^{\rm FOPT} =  0.3169 \,\pm\, 0.0077$. The average of Eq.~(\ref{eq:tOPEavg}) would then read $\alpha_s(m_\tau^2) =   0.3183 \pm0.0079$.}
\begin{equation}\label{eq:tOPEavg}
\alpha_s(m_\tau^2) \, = \, 0.3183 \,\pm\, 0.0069 \qquad\mbox{(tOPE, RF GC, ALEPH $V+A$)}.
\end{equation}
We adopt this averaging prescription in the RF GC analysis in the next section as well. The uncertainty of the average reflects the potential prospect of strong coupling determinations from hadronic $\tau$ spectral function moments in a situation where  an uncertainty in the parametrization of the nonperturbative corrections in the tOPE method would be absent. In Fig.~\ref{fig:Params} a visual comparison of the results for $\alpha_s(m_\tau^2)$ in the $\overline{\text{MS}}$ (left panel, see Eq.~(\ref{eq:alphastOPEMSbar})) and in the RF GC scheme (right panel, see Eq.~(\ref{eq:alphastOPERF})) is displayed. The average of Eq.~(\ref{eq:tOPEavg}) is also shown in the right panel.

We note that in Ref.~\cite{Pich:2016bdg} a number of other moment analyses using the CIPT and FOPT expansions have been carried out. We have checked that the observations described above for the improvement concerning the consistency of the extracted values for $\alpha_s(m_\tau^2)$ from the CIPT and FOPT expansions in the RF GC scheme are exemplary for all analyses carried out by them and not specific for the moment analysis in their Sec.~5.3.

\subsection[Single Weigth Function Analysis with Multiple   \texorpdfstring{$s_0$}{s0} values]{Single Weigth Function Analysis with Multiple \boldmath  \texorpdfstring{$s_0$}{s0} values}
\label{sec:singlemoment}

We turn now to the analysis by Boito et al.\ in Ref.~\cite{Boito:2020xli}.
This analysis follows the so-called "DV-model strategy" advocated in Refs.~\cite{Boito:2011qt} and used in previous analyses of the ALEPH and OPAL data in Refs.~\cite{Boito:2014sta,Boito:2012cr}.
In order to avoid the truncation of the OPE, one chooses weight functions with less (or even no) pinching, since this allows to employ low-degree polynomial weight functions $w(x)$ which strongly suppress higher-dimension OPE  corrections.  In the approximation of Eq.~(\ref{sec:DOPEnaive}) this leads to the absence of OPE corrections coming from higher-dimensional OPE condensates. However, this choice in general enhances the contributions  from DV effects, which  therefore must be included in the fit. This is done with  the parametrization of Eq.~(\ref{eq:DVparametrization}). Since DVs are related to residual resonance effects, the parameters $\delta_{V/A}$, $\gamma_{V/A}$,  $a_{V/A}$, and $b_{V/A}$ are channel dependent and have to be extracted from data.
The fits can be carried out using moments for the same weight function using different values for $s_0\le m_\tau^2$ under the assumption that the ansatz for the DV contribution in Eq.~(\ref{eq:DVparametrization}) is adequate.

In the analysis of  Ref.~\cite{Boito:2020xli}, the $\alpha_s$ results are based on the $V$ spectral function,
for which a more precise and updated data set was provided in the same reference, which includes information from recent  $e^+e^-\to {\rm hadrons}$ cross-section data related by isospin symmetry.
For the main results, moments for the weight function $w(x)=1$ are considered for  different values of $s_0$.  For this weight function all OPE corrections are strongly suppressed (and vanish in the approximation of Eq.~(\ref{sec:DOPEnaive})). As for the analysis of Ref.~\cite{Pich:2016bdg} discussed in the previous section, the approximation of Eq.~(\ref{sec:DOPEnaive}) is used so that no OPE corrections are included.
Using moments with several $s_0$ values simultaneously,  one can extract $\alpha_s$ together with the DV parameters from the fit. For the main results of Ref.~\cite{Boito:2020xli}, between fourteen and twenty $s_0$ values are included in each fit, ranging from $1.55$~GeV$^2$ to $m_\tau^2$, (see the entries in the last four lines in their table~1). In our analysis below we consider the fit with twenty $s_0$ values, starting at $s_0=1.55$~GeV$^2$. There are five fit parameters: $\alpha_s$ and the four parameters of the DV contribution. Other moments, including pinched moments requiring OPE condensate corrections, were also considered in Ref.~\cite{Boito:2020xli} as a consistency check of the analysis.

\begin{table}
\begin{center}
\begin{tabular}{|p{1.3cm}|p{1.5cm}|p{1.5cm}|p{1.5cm}|p{1.5cm}|p{1.5cm}|p{1.5cm}|p{1.3cm}|}
\hline
& central value&$\sigma_{\text{exp}}$&$\sigma_{\mu}$&$\sigma_{R}$&$\sigma_{N_g}$&$\sigma_5$&$\sigma_{\text{total}}$\\
\hline
 \multicolumn{8}{|c|}{ CIPT, $\chi^2$/dof = 12.97/15, $p$-value = 0.60} \\
 \hline
$\alpha_s(m_{\tau})$&0.3256&0.0089&0.0033&$-$&$-$&0.0021&0.0097\\
$\delta_V$&3.33&0.28&0.021&$-$&$-$&0.0055&0.28\\
$\gamma_V$&0.66&0.18&0.010&$-$&$-$&0.0029&0.18\\
$a_V$&$-$1.27&0.49&0.0026&$-$&$-$&0.0011&0.49\\
$b_V$&3.76&0.26&0.0023&$-$&$-$&0.0001&0.26\\
\hline
 \multicolumn{8}{|c|}{ FOPT, $\chi^2$/dof = 12.55/15, $p$-value = 0.64} \\
 \hline
$\alpha_s(m_{\tau})$&0.3083&0.0066&0.0014&$-$&$-$&0.0025&0.0072\\
$\delta_V$&3.51&0.28&0.054&$-$&$-$&0.027&0.29\\
$\gamma_V$&0.57&0.17&0.029&$-$&$-$&0.013&0.18\\
$a_V$&$-1.26$&0.48&0.022&$-$&$-$&0.0002&0.48\\
$b_V$&3.77&0.26&0.0063&$-$&$-$&0.0020&0.26\\
\hline
 \multicolumn{8}{|c|}{RF GC scheme, CIPT, $\chi^2$/dof = 12.70/15, $p$-value = 0.63} \\
\hline
$\alpha_s(m_{\tau})$&0.3159&0.0080&0.0033&0.0015&0.0035&0.0017&0.0096\\
$\delta_V$&3.44&0.28&0.028&0.055&0.041&0.0024&0.29\\
$\gamma_V$&0.61&0.17&0.013&0.027&0.020&0.0012&0.18\\
$a_V$&$-1.26$&0.49&0.0037&0.0023&0.0023&0.0007&0.49\\
$b_V$&3.77&0.26&0.0031&0.0054&0.0023&0.0001&0.26\\
\hline
 \multicolumn{8}{|c|}{RF GC scheme, FOPT, $\chi^2$/dof = 12.53/15, $p$-value = 0.64} \\
 \hline
$\alpha_s(m_{\tau})$&0.3081&0.0065&0.0015&0&0.0001&0.0025&0.0072\\
$\delta_V$&3.52&0.28&0.053&0.0037&0.0038&0.026&0.29\\
$\gamma_V$&0.57&0.17&0.029&0.0016&0.0018&0.013&0.18\\
$a_V$&$-1.26$&0.48&0.024&0.0018&0.0008&0.0003&0.48\\
$b_V$&3.77&0.26&0.0076&0.0010&0.0006&0.0019&0.26\\
\hline
\end{tabular}
\caption{Fitted parameters from the improved $\tau$ vector-isovector spectral function  of Ref.~\cite{Boito:2020xli} for CIPT and FOPT in the $\overline{\text{MS}}$ GC scheme and in the RF GC scheme. For the latter case  we use $N_g=0.57\pm0.23$. The uncertainty $\sigma_{c_{5,1}}$ reflects the variation of the 6-loop  coefficient as in Eq.~(\ref{eq:c51C}), while $\sigma_\mu$ and $\sigma_R$ refer to variations of the strong-coupling renormalization scale $\mu$ and  of the  IR subtraction scale $R$ as described in the text. The values for  $\sigma_\mu$, $\sigma_{5}$, $\sigma_R$, and
$\sigma_{N_g}$ are half the difference between the maximum and minimum values obtained from the $\xi$,
$c_{5,1}$, $R$, and $N_g$ variations, respectively, using the central values for the remaining parameters.  The parameters $b_V$ and $\gamma_V$ are given in units of GeV$^{-2}$; $a_V$ and $b_V$ are denoted  $\alpha_V$ and  $\beta_V$ in Ref.~\cite{Boito:2020xli}.}
\label{tab:DVResults}
\end{center}
\end{table}

The results of the fits based on the CIPT and FOPT expansions in the usual $\overline{\text{MS}}$ scheme  for the GC and using 20 values for $s_0$ are shown in the upper half of Tab.~\ref{tab:DVResults}.  The central values are again obtained using $\bar c_{5,1}=345$ for the 6-loop coefficient. For the strong coupling the results are
\begin{equation}
\label{eq:alphasDVsMSbar}
\begin{array}{l}
\alpha_s(m_\tau^2)^{\rm CIPT} \, = \, 0.3256 \,\pm\, 0.0097 \\
\alpha_s(m_\tau^2)^{\rm FOPT} \, = \, 0.3083 \,\pm\, 0.0072
\end{array}\, \qquad\mbox{(DV mod.,  $\overline{\rm MS}$ GC, new $V$ spec. func.)}.
\end{equation}
The experimental uncertainties $\sigma_{\rm exp}$ and the uncertainty $\sigma_{5}$, from the estimate of the 6-loop coefficient $\bar{c}_{5,1}$, are obtained in the same way as in our tOPE analysis. The uncertainty $\sigma_\mu$, arising from renormalization scale variations are obtained in an analogous way as well, but a lower bound is imposed on the $\xi$ variations for small $s_0$ values to avoid the appearance of nonperturbative scales. As the strict lower bound  for strong coupling renormalization scale variations we use $\mu_{\rm min}=0.7 m_\tau$ to safely stay outside the region of nonperturbative scales.  A similar prescription was applied in Ref.~\cite{Boito:2020xli}.

Our results reproduce very well those quoted in Eq. (4.1) and  Eq. (4.2)
of Ref.~\cite{Boito:2020xli}, for FOPT and CIPT, respectively.
The discrepancy between the results for $\alpha_s(m_\tau^2)$ from FOPT and CIPT is of $0.017$, which is 2 to 3 times larger than the  experimental uncertainties $\sigma_{\rm exp}$ which are about twice the size as in the tOPE analysis. The uncertainty from the renormalization scale variation $\sigma_{\mu}$ is significantly smaller than the experimental uncertainty and also smaller than for the tOPE analysis. As for the tOPE analysis, the CIPT-FOPT discrepancy in $\alpha_s(m_\tau^2)$ is much larger than the total uncertainty for both expansions which is obtained by adding all individual uncertainties quadratically. Interestingly, the fit results for the DV parameters for the CIPT and FOPT analysis are perfectly compatible within uncertainties indicating again that the asymptotic separation mostly affects the value of the strong coupling.

In the lower half of Tab.~\ref{tab:DVResults}, we show the results of the fits using the CIPT and FOPT expansions in the RF GC scheme.
The central values are obtained from using $N_g=0.57$ and $R=0.8\, \sqrt{s_0}$ as well as $\bar c_{5,1}=345$. For the strong coupling the results are
\begin{equation}
\label{eq:alphasDVsRF}
\begin{array}{l}
\alpha_s(m_\tau^2)^{\rm CIPT} \, = \, 0.3159 \,\pm\, 0.0096 \\
\alpha_s(m_\tau^2)^{\rm FOPT} \, = \, 0.3081 \,\pm\, 0.0072
\end{array}\, \qquad\mbox{(DV mod.,  RF GC, new $V$ spec. func.)}.
\end{equation}
The uncertainties from the truncation of the perturbation series, $\sigma_\mu$ and $\sigma_{c_{5,1}}$ are estimated as in the $\overline{\rm MS}$ GC analysis. The uncertainty from the IR factorization scale $R$ is based on the variation range $0.7\sqrt{s_0}\leq R \leq \sqrt{s_0}$, where for smaller values of $s_0$ the absolute lower bound $R_{\rm min}=0.7m_\tau$ is imposed on $R$, in analogy to the renormalization scale. As for the tOPE analysis, the FOPT results are virtually unchanged and the uncertainties $\sigma_R$ and $\sigma_{N_g}$ are negligibly small. This again reflects the compatibility of the FOPT expansion with the standard form of the OPE.

The CIPT results, on the other hand, change again significantly. The difference of the CIPT and FOPT central values for $\alpha_s(m_\tau^2)$ is reduced to 0.008. This is more than twice as the difference for the tOPE analysis in the RF GC scheme, but still half of the difference that is obtained when the $\overline{\rm MS}$ GC scheme is employed. Overall, there is  a much better agreement for the $\alpha_s(m_\tau^2)$ determinations in the RF GC scheme. There are, however, a few notable differences. While in the tOPE analysis the uncertainty in $\alpha_s(m_\tau^2)$ increased moderately due to $\sigma_R$ and $\sigma_{N_g}$, here these two uncertainties are still relatively small and do not lead to any noticeable increase in the total uncertainty. The total uncertainty for the CIPT results even slightly decreases as a result of a smaller experimental uncertainty $\sigma_{\text{exp}}$. Furthermore, while in the tOPE analysis
$\sigma_R$ and $\sigma_{N_g}$ are significant for the CIPT expansion and there are notable changes in the central values of the OPE condensates, here the impact of $\sigma_R$ and $\sigma_{N_g}$ is smaller in comparison and the fit results for the DV parameters change very little. Finally, while for the tOPE analysis the $p$-value decreased in the RF GC scheme, no modification of the $p$-value is observed here.

\begin{figure}
	\centering
		\includegraphics[width=\textwidth]{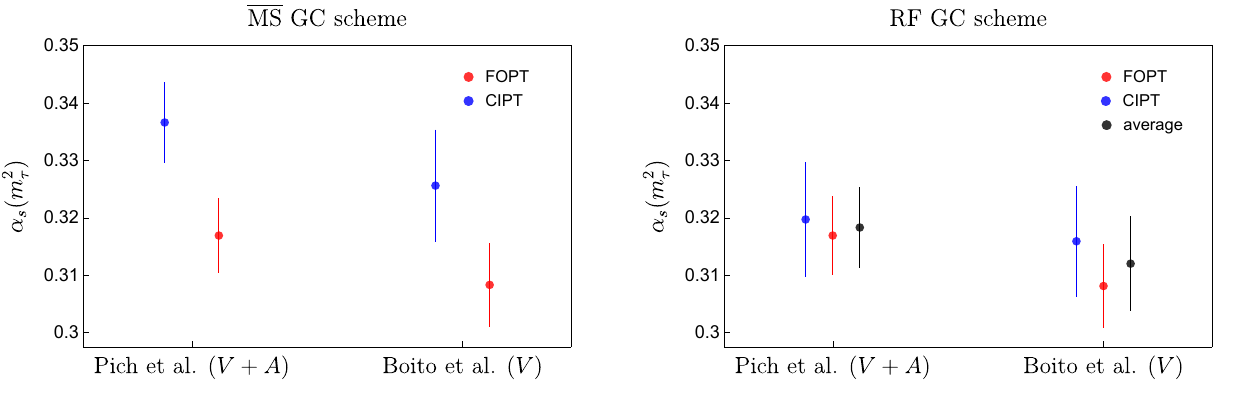}
	\caption{\label{fig:Params}
	Left panel: Results for $\alpha_s(m_\tau^2)$ in FOPT (red) and CIPT (blue) in the $\overline{\rm MS}$ GC scheme, following the strategies  of Ref.~\cite{Pich:2016bdg} (Pich et al.), based on the tOPE strategy applied to the $V+A$ ALEPH data~\cite{Davier:2013sfa}, and Ref.~\cite{Boito:2020xli} (Boito et al.),  based on the DV-model strategy applied to the new vector spectral function of Ref.~\cite{Boito:2020xli}. Right panel: Results for $\alpha_s(m_\tau^2)$ in FOPT (red) and CIPT (blue) and their average (black) in the RF GC scheme for the same analysis set-ups. The substantial reduction in the discrepancy between FOPT and CIPT result is clearly evident.}
\end{figure}

With the strong suppression of the asymptotic separation in  the RF GC scheme, it becomes possible to average the FOPT and CIPT results of Eq.~(\ref{eq:alphasDVsRF}). Following the prescription described in the previous section we find\footnote{If we used $\pm 400$ for the uncertainty in $\bar c_{5,1}$, as in Ref.~\cite{Pich:2016bdg},  $\sigma_5$ would increase to $0.0059$ for CIPT and $0.0074$ for FOPT in the $\overline{\rm MS}$ GC scheme, which would yield
$\alpha_s(m_\tau^2)^{\rm CIPT} =  0.326 \,\pm\, 0.011$ and
$\alpha_s(m_\tau^2)^{\rm FOPT} =  0.308 \,\pm\, 0.010$.
%If we had adopted uncertainties of $\pm 400$ for $\bar c_{5,1}$, $\sigma_5$ would increase to $0.0049$ and CIPT and $0.0039$ for FOPT, which would yield
In the RF GC scheme, a similar increase in uncertainty is observed and the final values would be $\alpha_s(m_\tau^2)^{\rm CIPT} =  0.316 \,\pm\, 0.011$ and  $\alpha_s(m_\tau^2)^{\rm F	OPT} =  0.308 \,\pm\, 0.010$. The average of Eq.~(\ref{eq:DVavg}) would  read $\alpha_s(m_\tau^2) =   0.312 \pm 0.011$.}
\begin{equation}
\label{eq:DVavg}
	\alpha_s(m_\tau^2) \, = \, 0.3120 \,\pm\, 0.0082 \qquad\mbox{(DV mod.,  RF GC, new $V$ spec. func.)}.
\end{equation}
This result reflects the prospect  for strong coupling determinations under the assumptions made in the DV-model strategy.
In Fig.~\ref{fig:Params} a visual comparison of the results for $\alpha_s(m_\tau^2)$ in the $\overline{\text{MS}}$ (left panel, see Eq.~(\ref{eq:alphasDVsMSbar})) and in the RF GC scheme (right panel, see Eq.~(\ref{eq:alphasDVsRF})) is displayed. The average of Eq.~(\ref{eq:DVavg}) is also shown in the right panel.

As already mentioned above, in Ref.~\cite{Boito:2020xli} the same type of fits have also been carried out for smaller number of $s_0$ values, and also some analyses based on a combination of different types moments were carried out. We emphasize that the observations concerning the good agreement of the CIPT and FOPT determinations of $\alpha_s(m_\tau^2)$ is general and not dependent on the particular analysis set up described above. We also note that it may appear tempting to combine the tOPE average of Eq.~(\ref{eq:tOPEavg}) with the DV model average in Eq.~(\ref{eq:DVavg}) taking their difference, which only amounts to $0.0063$, as an estimate for the treatment of nonperturbative effects. We refrain from such a treatment since the $V$ spectral function determined in Ref.~\cite{Boito:2020xli} has not yet been analysed in the tOPE approach.
Furthermore, it is well known from the experience with the ALEPH data that the tOPE and the DV-model strategies can lead to rather discrepant $\alpha_s$ results when the same data set is analysed~\cite{Boito:2014sta,Boito:2016oam,Pich:2016bdg}. This discrepancy should be attributed to the treatment of nonperturbative effects, and a detailed investigation is beyond the scope of this work.

\vspace{1cm}

\section{Conclusions}
\label{sec:conclusions}

In this work, which is Part~II of a series of articles, we have applied the renormalon-free scheme for the gluon condensate (GC) defined in Part I~\cite{Benitez-Rathgeb:2022yqb} to the FOPT and CIPT perturbative expansions for $\tau$ hadronic spectral function moments and examined it from the phenomenological perspective. The scheme is based on a perturbative redefinition of the GC matrix clement in close analogy to the well-known implementation of short-distance heavy quark mass schemes, where the scheme changes induce perturbative order-by-order subtractions to the original perturbative coefficients. The renormalon-free scheme, which was discussed in detail in Part~I and which we call the RF gluon condensate scheme, depends on an IR factorization scale $R$ and the normalization $N_g$ of the GC renormalon. The RF GC matrix element is scale-invariant and can be easily related to other schemes  through its perturbative definition (including the non-renormalon-free original $\overline{\rm MS}$ scheme). In addition to the removal of renormalon related divergent perturbative contributions in the series, which is well-known from the use of short-distance quark mass schemes, the RF GC scheme also reconciles the long-standing discrepancy between the CIPT and FOPT expansions for the perturbative series of $\tau$ hadronic spectral function moments, which appeared for spectral function moments where the GC OPE correction is strongly suppressed (and which we call GC suppressed). As was shown by Hoang and Regner~\cite{Hoang:2020mkw,Hoang:2021nlz}, the CIPT expansion for such moments still has a strong quartic (and higher power) sensitivity to IR moments and is therefore not compatible with the standard analytic form for OPE corrections. This effect, which is, numerically, strongly dominated by the GC renormalon, is much larger than the size of the GC OPE correction itself and thus renders the CIPT expansion inconsistent if the usual $\overline{\rm MS}$ scheme is employed for the OPE condensate matrix elements. By switching to the RF GC scheme, this inconsistency is reduced to a negligible level, such that the CIPT expansion is practically cured. While switching to the RF GC scheme for such moments has very little effects for the FOPT expansion, the CIPT expansion is modified substantially such that the discrepancy between them is resolved.

In this article we explored the impact of the uncertainties of the GC renormalon norm $N_g$ and of variations of the IR subtraction scale $R$ on the CIPT expansion in the RF gluon condensate scheme for GC suppressed moments. Analysing three different methods, which all yield consistent results, we determined $N_g^{(n_f=3)}=0.57\pm 0.23$, having a relative uncertainty of $40\%$. We showed that this uncertainty and variations of the IR subtraction scale $R$ increase the perturbative uncertainties of the CIPT expansion, but we found that
this increase is by far outweighed by the improved consistency between the CIPT and FOPT expansion.

We have demonstrated the improved consistency in the RF GC scheme by applying the CIPT and FOPT expansions in the context of two full-fledged state-of-the art $\alpha_s(m_\tau^2)$ determination approaches from the recent literature by Pich and Rodr\'iguez-Sanchez~\cite{Pich:2016bdg} and by Boito, Golterman, Maltman, Peris, Rodrigues and Schaaf~\cite{Boito:2020xli} and accounting for the additional uncertainties related to the RF GC scheme. These two references are representatives of the two major approaches concerning the treatment of nonperturbative corrections currently used in the literature, called the truncated OPE approach and the duality-violation-model strategy. For both approaches we find a substantially improved consistency between the FOPT and CIPT expansions and that the uncertainties due to $N_g$ and $R$, that mostly affect the CIPT expansion, are quite small. Since the CIPT expansion has quite different perturbative properties than the FOPT expansion related to the all-order resummations of phase corrections~\cite{Pivovarov:1991rh}, it thus remains a very valuable method that should still be employed in future phenomenological analysis -- but only if it is employed within a renormalon-free GC scheme.

We believe that -- for all practical matters -- the long-standing CIPT-FOPT discrepancy problem can now be considered as resolved.
This resolution depends on the acceptance of the proposition that the values of the known Adler function perturbative coefficients up to ${\cal O}(\alpha_s^4)$, which are perfectly consistent with the sizeable GC norm value $N_g^{(n_f=3)}=0.57\pm 0.23$, are indeed affected by the GC renormalon and not created by some accidental unrelated finite-order behavior. This proposition can, as a matter of principle, not be proven. But the same type of proposition is used in all phenomenological applications where perturbative subtractions of renormalons are important, most notably in heavy quark physics. We therefore believe that it is reasonable to accept this proposition for future analyses where the GC or the GC renormalon play important roles. As far as the analyses of $\tau$ hadronic spectral function moments are concerned, aspects which so far may have been considered subleading, such as uncertainties related to the treatment of nonperturbative corrections or the treatment of different data sets~\cite{ParticleDataGroup:2020ssz,dEnterria:2022hzv}, can now receive undivided attention.

\section*{Acknowledgments} DB and MJ would like to thank the Particle Physics Group of the University of Vienna for hospitality.
We acknowledge partial support by the FWF Austrian Science Fund under the Doctoral Program ``Particles and Interactions'' No.\ W1252-N27 and under the Project No. P32383-N27.  We also thank the Erwin-Schr\"odinger International Institute for Mathematics and Physics for partial support. DB's work was supported by   by the
S\~ao Paulo Research Foundation (FAPESP) Grant No. 2021/06756-6, by CNPq Grant
No. 308979/2021-4, and by Coordena\c c\~ao de Aperfei\c coamento de Pessoal de N\'ivel Superior
– Brasil (CAPES) – Finance Code 001.
\vspace*{0.3cm}

\begin{appendix}

\section{GC Renormalon Norm Conventions}
\label{app:normconventions}

In this appendix we compare our norm convention for the renormalon calculus with the conventions used in \cite{Beneke:2008ad,Lee:2011te,Bali:2014fea} and provide conversion formulae. All expressions below are written down using
\begin{equation}
\label{eq:betafct}
\frac{d\alpha_s(Q^2)}{d\ln Q}\, = \, \beta(\alpha_s(Q^2)) \, \equiv \,
-2\,\alpha_s(Q^2)\,\sum\limits_{n=0}^\infty \beta_n\Big(\frac{\alpha_s(Q^2)}{4\pi}\Big)^{n+1}
\end{equation}
as the definition of the coefficients of the QCD $\beta$-function such that we have $\beta_0=11-2\,n_f/3$ and $\beta_1=102-38\,n_f/3$ for the one- and two-loop coefficients for $n_f$ dynamical flavors. We also adopt the abbreviation $a\equiv\alpha_s\beta_0/(4\pi)$.

In our convention we write the perturbation series for a generic quantity as
\begin{eqnarray}
\label{eq:oursigma}
\sigma \, = \, \sum_{n=1}^\infty\, c_n\,a^n\,.
\end{eqnarray}
The corresponding Borel function is defined as
\begin{eqnarray}
\label{eq:ourB}
B[\sigma](u) \, \equiv \, \sum_{n=1}^\infty\, \frac{c_n}{\Gamma(n)}\,u^{n-1}\,,
\end{eqnarray}
which gives
\begin{equation}
\label{eq:ourBorel}
\sigma = \int_0^\infty \!\! {\rm d} u \,
B[\sigma](u)\,e^{-\frac{u}{a}}
\end{equation}
for the inverse Borel integration that gives back the original series $\sigma$. The generic form that is adopted for a non-analytic IR renormalon term in the Borel function related to a dimension $d=2p$ OPE correction reads
\begin{eqnarray}
\label{eq:ourBfct}
B[\sigma](u) & \sim & \frac{N_{2p}}{(p-u)^{\gamma}}\,,
\end{eqnarray}
which has a branch point at $u=p$.

In Ref.~\cite{Beneke:2008ad} (Beneke and Jamin) the perturbation series for the generic quantity $\sigma$ is written as
\begin{eqnarray}
\label{eq:BJsigma}
\sigma \, = \, \sum_{n=0}^\infty\, p_n\,\alpha_s^{n+1}\,,
\end{eqnarray}
and the Borel function is defined as
\begin{eqnarray}
\label{eq:BJB}
\hat B[\sigma](t) \, \equiv \, \sum_{n=0}^\infty\, \frac{p_n}{\Gamma(n+1)}\,t^{n}\,.
\end{eqnarray}
The inverse Borel integration has the form
\begin{equation}
\label{eq:BJBorel}
\sigma = \int_0^\infty \!\! {\rm d} t \,
\hat B[\sigma](t)\,e^{-\frac{t}{\alpha_s}}
\, = \
{\textstyle \frac{4\pi}{\beta_0}}\, \int_0^\infty \!\! {\rm d} u \,
 \hat B[\sigma]\Big({\textstyle\frac{4\pi}{\beta_0}u}\Big)\,e^{-\frac{u}{a}}
\,,
\end{equation}
and the generic form adopted for a non-analytic IR renormalon term in the Borel function is
\begin{eqnarray}
\label{eq:BJBfct}
\hat B[\sigma]\Big({\textstyle\frac{4\pi}{\beta_0}u}\Big) & \sim & \frac{d_p^{\rm IR}}{(p-u)^{\gamma}}\,.
\end{eqnarray}

In Refs.~\cite{Lee:2011te} (Lee) and \cite{Bali:2014fea} (Bali~et~al.) the perturbation series for the generic quantity $\sigma$ is written as
in Eq.~(\ref{eq:BJsigma}) and the Borel function is defined as
\begin{eqnarray}
\label{eq:LeeB}
\tilde B[\sigma](u)\,=\, \tilde \sigma(u)\,=\, \sum_{n=0}^\infty\, \frac{p_n}{\Gamma(n+1)}\,\frac{(4\pi)^n u^n}{\beta_0^n}\,.
\end{eqnarray}
The inverse Borel integration reads
\begin{equation}
\label{eq:LeeBorel}
\sigma =
{\textstyle \frac{4\pi}{\beta_0}}\, \int_0^\infty \!\! {\rm d} u \,
\tilde B[\sigma](u)\,e^{-\frac{u}{a}}
\,,
\end{equation}
and a generic non-analytic IR renormalon term in the Borel function is written as
\begin{eqnarray}
\label{eq:LeeBfct}
\tilde B[\sigma](u)=\tilde \sigma(u) & \sim & \frac{{\cal N}_p}{(1-u/p)^{\gamma}}\,.
\end{eqnarray}

The various Borel function definitions are related to ours through
\begin{eqnarray}
\label{eq:Brelations}
B[\sigma](u) \, = \,
{\textstyle \frac{4\pi}{\beta_0}}\,
\hat B[\sigma]\Big({\textstyle\frac{4\pi}{\beta_0}u}\Big) \, = \,
{\textstyle \frac{4\pi}{\beta_0}}\,
\tilde B[\sigma](u)\, = \,
{\textstyle \frac{4\pi}{\beta_0}}\,
\tilde\sigma(u)\,,
\end{eqnarray}
which leads to the following relation of the norms
\begin{eqnarray}
\label{eq:norms}
N_{2p}
\, = \,
{\textstyle \frac{4\pi}{\beta_0}} \,d_p^{\rm IR}
\, = \,
{\textstyle \frac{4\pi}{\beta_0}}\,p^\gamma \,{\cal N}_p\,.
\end{eqnarray}

The norm for the renormalon terms also depends on the scheme used for the strong coupling. For two strong coupling schemes, $a$ and $a^*$,  which are related by $a^*/a = 1 - \lambda \, a +\ldots\sim (1+\lambda\, a+\ldots)^{-1}$, their QCD scales are related by $\Lambda_{\rm QCD}=\Lambda_{\rm QCD}^*\,e^{\lambda/2}$. This implies that the Borel function in the two schemes satisfy $B[\sigma](u)=B^*[\sigma](u)\,e^{\lambda u}$, so that their respective IR renormalon norms are related by $N_{2p}=N_{2p}^*\,e^{p \lambda}$. For a UV renormalon term with a branch point at $u=k<0$ the analogous relation holds for $p$ replaced by $k$. For the relation between the $C$-scheme (for $C=0$)~\cite{Boito:2016pwf} and the $\overline{\rm MS}$ scheme we have $\lambda=0$ (see the appendix of Part~I~\cite{Benitez-Rathgeb:2022yqb}), so that their QCD scales and renormalon norms are identical.

\section{GC Renormalon Norm in the Quenched Approximation}
\label{sec:GCquenched}

In the quenched approximation the GC renormalon norm $N_g^{(n_f=0)}$ was determined previously in two dedicated analyses by Lee~\cite{Lee:2011te} and Bali~et~al.~\cite{Bali:2014fea}. Lee used the conformal mapping approach described in Sec.~\ref{sec:cmapproach} up to ${\cal O}(\alpha_s^4)$ employing the mapping of Eq.~(\ref{eq:mappingLee}). He obtained  ${\cal N}_{2}^{\hat D,(n_f=0)}=0.32 - 0.13 + 0.026 + 0.075  = 0.29$ for the GC norm of the Adler function. Accounting for the conventional factor $2\pi^2/3$ in the Adler function's GC OPE correction shown in Eq.~(\ref{eq:AdlerOPEGCv2}) and switching to our normalization convention this corresponds to $N_g^{(n_f=0)}=0.36 - 0.15 + 0.029 + 0.084 = 0.32$ for the GC renormalon norm.
Bali~et~al. used numerical stochastic perturbation theory to determine the perturbative series for the average plaquette $P_{\rm pert}(a)\equiv \langle P\rangle = \sum_{n=0} p^{\rm latt}_n \alpha^{n+1}(a)$ for the lattice spacing $a$ up to ${\cal O}(\alpha^{35})$ in the infinite volume limit based on the standard Wilson gauge action. Here $\alpha(a)$ stands for the strong coupling at lattice spacing $a$ which should not be confused with our abbreviation for the strong coupling in Eq.~(\ref{eq:adef}). Using a formula for the asymptotic large-$n$ behavior of the GC renormalon series $p_n^{\rm latt(asy)}$, the GC renormalon norm was determined from the ratio $p^{\rm latt}_n/p_n^{\rm latt(asy)}$. They obtained the result ${\cal N}_{2,{\rm latt}}^{P,(n_f=0)}=(42\pm 17)\times 10^4$ for the GC renormalon norm of the plaquette in the strong coupling lattice scheme, which results in ${\cal N}_{2,{\rm latt}}^{P,(n_f=0)}=(0.61\pm 0.25)$ in $\overline{\rm MS}$ strong coupling scheme. Accounting for the conventional factor $\pi^2/36$ in the OPE corrections of the plaquette and switching to our normalization convention this corresponds to $N_g^{(n_f=0)}=(16.4\pm 6.7)$ for the GC renormalon norm.
The results obtained by Lee and Bali~et~al.\ are clearly incompatible. It is therefore worth to apply the methods we have discussed in Sec.~\ref{sec:GCnormalization} in the quenched approximation. Since theoretical studies of the Adler function  in the quenched approximation have so far not relied on estimates of the ${\cal O}(\alpha_s^5)$ coefficient $\bar c_{5,1}$, we will in the following only use the known cofficients up to ${\cal O}(\alpha_s^4)$.

\begin{table}
	\begin{center}
		\begin{tabular}{ |p{3.5cm}|p{1.7cm}|   }
			\hline
			&$m=4$\\
			\hline
			$w(u,5)$&$0.88$\\
			$w(u,10)$&$0.76$\\
			$w(u,15)$&$0.73$\\
			\hline
			$\chi^2_{\xi=1} ~ (\sqrt{s_0}=m_{\tau})$&$0.86 \pm 0.28$\\
			$\chi^2_{\xi=2} ~ (\sqrt{s_0}=m_{\tau})$&$0.65 \pm 0.25$\\
			$\chi^2_{\xi=1} ~ (\sqrt{s_0}=3 ~ {\rm GeV})$&$0.64 \pm 0.22$\\
			$\chi^2_{\xi=2} ~ (\sqrt{s_0}=3 ~ {\rm GeV})$&$0.55 \pm 0.21$\\
			\hline
		\end{tabular}
		\caption{\label{tab:nf0confmapp} Results for $N_g$ for the $n_f=0$ flavor scheme at ${\cal O}(\alpha_s^4)$ using the conformal mapping approach (upper part) as well as the optimal subtraction approach (lower part).}
	\end{center}
\end{table}

\begin{figure}
	\centering
		\includegraphics[width=\textwidth]{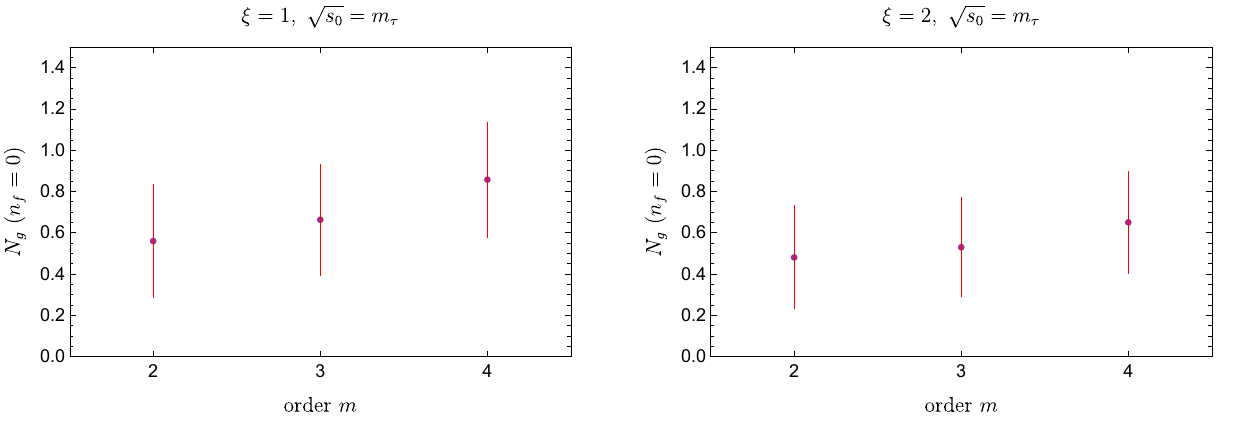}
	\caption{\label{fig:chi2nf0n4}
		Results for $N_g$ at ${\cal O}(\alpha_s^{2,3,4})$ using the optimal subtraction approach for $\xi=1$ (left panel) and $\xi=2$ (right panel) for the $n_f=0$ flavor scheme.}
\end{figure}

Let us start with the Borel function model approach. As was already described in Ref.~\cite{Beneke:2008ad}, one can use the Borel model of Eq.~(\ref{eq:Bmodel}) dropping the linear term $b^{(1)}u$ given that its coefficient is quite small. The smallness of $b^{(1)}$ obtained for our $n_f=3$ analysis in Sec.~\ref{sec:mr-model} indicates that the other terms contained in the Borel model should be sufficient to describe the known coefficients, so that dropping this term does not deteriorate the quality of the model. Using only the Adler function cofficients up to ${\cal O}(\alpha_s^4)$ this yields $N_g=0.63$ for $n_f=3$, which is almost identical to the case when term $b^{(1)}u$  and the estimate of $\bar c_{5,1}$ are included, see the numbers quoted below Eq.~(\ref{eq:Bmodel}). Applying the same method for $n_f=0$ ($\hat b_1=51/121=0.421$) in the $C$-scheme we obtain
\begin{align}
\label{eq:Bmodelnf0}
\begin{split}
B[\hat{D}(s)]_{\text{mr}}^{(0)}(u) &= b^{(0)} + \frac{2\pi^2}{3}\frac{N_g^{(n_f=0)}
	\left[1-\frac{152}{363}\bar{a}(-s)\right]}{(2-u)^{1+4\hat b_1^{(0)}}} + \frac{N_{6}^{(0)}}{(3-u)^{1+2\hat b_1^{(0)}}}+ \frac{N_{-2}^{(0)}}{(1+u)^{\gamma_{2-2\hat b_1^{(0)}}}}\,,
\end{split}
\end{align}
with $N_g^{(n_f=0)}=0.98$, $ N_6^{(0)}= -26.12$, $N_{-2}^{(0)}= 0.022$ and $b^{(0)}= -0.12$, where the coefficients of the Adler function are given by $\bar c_{1,1}^{(0)}=1$, $\bar c_{2,1}^{(0)}=1.986$, $\bar c_{3,1}^{(0)}=20.985$, $\bar c_{4,1}^{(0)}=161.224$.
Applying the conformal mapping approach for that same Borel function model, we again find that Lee's mapping function yields a very slowly converging series that undershoots the actual result significantly at ${\cal O}(\alpha_s^4)$. This indicates that his estimate $N_g^{(n_f=0)}= 0.32$ is somewhat low. The mapping functions of Eq.~(\ref{eq:mappingCaprini}) again provide much better results and yield the results shown in the upper part of Tab.~\ref{tab:nf0confmapp} for $p=5,10,15$.
Finally, we apply the optimal subtraction approach, where we can also obtain a reliable result using only information up to ${\cal O}(\alpha_s^4)$. Following the method as described in Sec.~\ref{sec:optimalsubtraction}, we find the results for $N_g$ shown in the lower part of Tab.~\ref{tab:nf0confmapp} at order $m=4$ for $\sqrt{s_0}=m_\tau$ and $3$~GeV and $\xi=1,2$. For illustration, in Fig.~\ref{fig:chi2nf0n4} we also display
$N_g$ obtained for $\sqrt{s_0}=m_\tau$ and $\xi=1,2$ at order $m=2,3,4$. Adopting the envelope of all order $m=4$ results, our final result for $N_g^{(n_f=0)}$ reads:
\begin{equation}
\label{eq:Ngoptsubnf0}
N_g^{(n_f=0)}=0.74\pm 0.40\,.
\end{equation}
The results we obtain from the Borel function model as well as from the conformal mapping approaches are fully compatible with Eq.~(\ref{eq:Ngoptsubnf0}), and also Lee's result is compatible within uncertainties. The result, however, disagrees with Bali~et al.
We note that the relative uncertainty in $N_g^{(n_f=0)}$ obtained by the optimal subtraction approach is $54\%$ and somewhat larger than that for the case $n_f=3$ in Eq.~(\ref{eq:Ngoptsub}). This is potentially related to the higher IR sensitivity of QCD perturbation theory in the quenched approximation.

In App.~\ref{sec:commentBalietal} we provide arguments that suggest that Bali~et al. may have underestimated the uncertainty in their result. We therefore consider the result quoted in Eq.~(\ref{eq:Ngoptsubnf0}) as a reliable determination of the GC renormalon norm in the quenched approximation.
We note that a scenario where the GC renormalon norm $N_g^{(n_f=0)}$ would be around $16$ corresponds to a situation that strongly contradicts the naturalness assumption we discussed in Sec.~\ref{sec:GCnormalization} and the proposition that the Adler function series at ${\cal O}(\alpha_s^4)$ is already largely determined by the terms accounted for in the Borel function model of Eq.~(\ref{eq:Bmodel}). In fact, the parameters of the Borel function in such a scenario would have to be highly fine-tuned, so that the large size of the $N_g^{(n_f=0)}$ is hidden in the perturbative coefficients up to ${\cal O}(\alpha_s^4)$. By constructing and analyzing a number of Borel function models with such a large value for $N_g^{(n_f=0)}$, we have found that in such a scenario the perturbation series for GCS spectral function moments entirely change their nicely converging character beyond ${\cal O}(\alpha_s^4)$, so that the values the moments series approach for orders up to ${\cal O}(\alpha_s^4)$ are far away from the values the truncated moment series approach at higher orders. In other words, the truncated series values at ${\cal O}(\alpha_s^4)$ including the estimate of the truncation error would be far away from the true value of the series.

\section{Comment on the Gluon Condensate Norm from Lattice QCD}
\label{sec:commentBalietal}

\begin{figure}
	\centering
		\includegraphics[width=\textwidth]{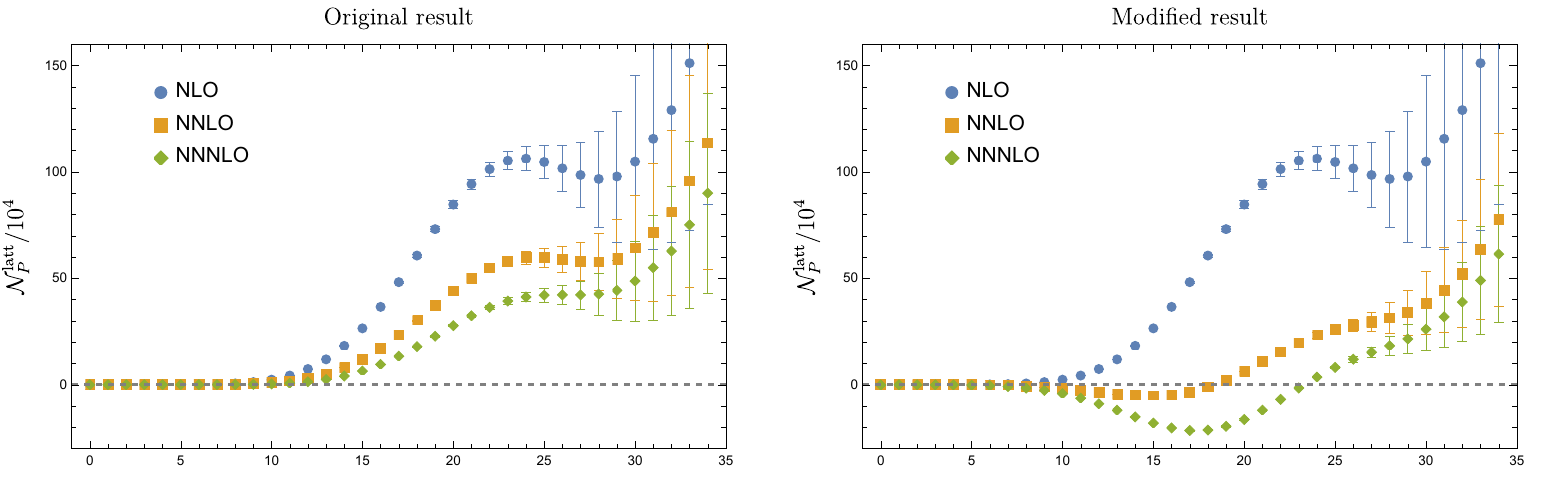}
	\caption{\label{fig:pinedaplots}
		Left panel: Results as a function of order for the GC renormalon norm of the average plaquette obtained in Ref.~\cite{Bali:2014fea}. Right panel: The corresponding results when the $1/n$ subleading asymptotic corrections are consistently expanded. In blue we show results at NLO, in orange at NNLO and in green at NNNLO.}
\end{figure}

In Ref.~\cite{Bali:2014fea} Bali~et~al.\ determined the normalization of the GC renormalon  using SU(3) lattice QCD in the quenched approximation ($n_f=0$). They used numerical stochastic perturbation theory to determine the perturbative series for the average plaquette $P_{\rm pert}(a)\equiv \langle P\rangle = \sum_{n=0} p^{\rm latt}_n \alpha^{n+1}(a)$ for the lattice spacing $a$ up to ${\cal O}(\alpha^{35})$ in the infinite volume limit based on the standard Wilson gauge action.  Using the known formula for the asymptotic behavior of the GC renormalon series
\begin{align}
	\begin{split}
		p_n^{\rm latt(asy)} &\stackrel{n\to\infty}{=} {\cal N}_P\, \Big(\frac{\beta_0}{8\pi}\Big)^n\,\frac{\Gamma(n+1+4\hat b_1)}{\Gamma(1+4\hat b_1)}\, \times
		\\
		&\hspace{3.0cm}\times \left\{1+ \frac{20.08931}{n+4\hat{b}_1} + \frac{505 \pm 33}{(n+4\hat{b}_1)(n+4\hat{b}_1-1)} + \mathcal{O}\left( \frac{1}{n^3} \right) \right\}
	\end{split}
	\label{eq:pnlattasy}
\end{align}
the plaquette's GC renormalon norm ${\cal N}_P$ was determined from the ratio $p_n^{\rm latt}/p_n^{\rm latt(asy)}$ using that the GC renormalon is the renormalon located closest to the origin in the Borel plane. At the hadron level the average plaquette for the lattice spacing $a$ has the form
\begin{align}
	\label{eq:Phad}
	P(a) = P_{\rm pert}(a) + \frac{\pi^2}{36}[1+{\cal O}(\alpha(a))]\, a^4\,\langle\bar G^2\rangle
\end{align}
accounting for the GC OPE correction. So the norm of the GC renormalon associated to the GC matrix element $\langle\bar G^2\rangle$ is $36/\pi^2$ times the GC renormalon normalization of the plaquette.
Bali~et~al.\ argued that at orders around $n=26$  the large-order asymptotics of the GC renormalon saturates  the coefficients $p_n^{\rm latt}$ and the uncertainties from lattice perturbation theory are still sufficiently small, such that a reliable value for ${\cal N}_P$ can be determined. The result for the ratio as determined in Ref.~\cite{Bali:2014fea} is shown in the left panel of Fig.~\ref{fig:pinedaplots}, where the error bars represent the uncertainties in the lattice coefficients $p_n^{\rm latt}$. The colored symbols represent the results obtained from the formula in Eq.~(\ref{eq:pnlattasy}) using the different approximations concerning the $1/n$ corrections in the asymptotic large-$n$ limit, where `NLO' stands for the dominant term without $1/n$ corrections\footnote{The terminology `NLO' stems from the fact that the leading asymptotic contribution already involves the term $\hat b_1$ which contains the 2-loop $\beta$-function coefficient $\beta_1$.}, `NNLO' stands for including the term $20.08931/(n+4\hat b_1)$ and so on.

The lattice spacing also governs the UV renormalization in lattice perturbation theory and entails a particular scheme for the strong coupling. Since the known 3- and 4-loop coefficients $\beta_{2,3}^{\rm latt}$ of the lattice scheme $\beta$-function\footnote{The 4-loop coefficient $\beta_3^{\rm latt}$ is only known numerically with a sizeable uncertainty. This uncertainty causes the error in the numerator of the NNNLO term in Eq.~(\ref{eq:pnlattasy}).} are extremely large (see Sec.~III~A in Ref.~\cite{Bali:2014fea}), the asymptotic $1/n$ corrections in Eq.~(\ref{eq:pnlattasy}) are very large as well, so that the expression for the asymptotic behavior of the coefficients $p_n^{\rm latt(asy)}/{\cal N}_P$ itself has large theoretical uncertainties even for orders where the GC renormalon may completely saturate the perturbative coefficients.
For order $n=26$ (and $n_f=0$) the series of subleading asymptotic terms in the curly brackets of Eq.~(\ref{eq:pnlattasy}) reads $\{1~(\mbox{NLO}) + 0.725614~(\mbox{NNLO}) + 0.683517~(\mbox{NNNLO})+{\cal O}(1/n^3)\}$. We see that the convergence of the series in the curly brackets is quite bad
so that the value of $p_{26}^{\rm latt(asy)}$ has a large uncertainty.\footnote{\label{ftnt:MSbar} In contrast, in the common $\overline{\rm MS}$ scheme for the strong coupling the corresponding series of subleading terms at order $n=26$ reads $\{1~(\mbox{NLO}) -0.11329~(\mbox{NNLO}) -0.0014895~(\mbox{NNLO})+{\cal O}(1/n^3)\}$~\cite{Bali:2014fea}. Here the convergence is excellent.}
Bali~et al.\ stated that ${\cal N}_P$ can be extracted at order $n=26$  from the (green) NNNLO result using the difference between NNLO and NNNLO as an estimate for that uncertainty. They obtained ${\cal N}_P=(42\pm 17)\times 10^4$, which corresponds to
$N_g^{(n_f=0)} = 16.43 \pm 6.73$
for the GC renormalon normalization in our convention.

A possible alternative way to calculate the ratio $p_n^{\rm latt}/p_n^{\rm latt(asy)}$ and to test for the impact of the uncertainties related to the sizeable  subleading asymptotic terms in the lattice strong coupling scheme is to systematically expand in the subleading $1/n$ corrections in the ratio $p_n/p_n^{\rm latt(asy)}$:
\begin{align}
	\begin{split}
		p_n^{\rm latt}/p_n^{\rm latt(asy)} &= \frac{p_n^{\rm latt}}{{\cal N}_P}   \Big(\frac{8\pi}{\beta_0}\Big)^n\,\frac{\Gamma(1+4\hat b_1)}{\Gamma(n+1+4\hat b_1)} \times
		\\
		&\times\left\{1-\frac{20.08931}{n+4 \hat{b}_1}-\frac{505 \pm 33}{(n+4 \hat{b}_1)(n+4 \hat{b}_1-1)}+\frac{404}{(n+4 \hat{b}_1)^2} + \mathcal{O}\left( \frac{1}{n^3} \right)  \right\} .
	\end{split}
	\label{eq:asyratio}
\end{align}
If the corrections were well under control, the outcome based on this formula would be equivalent to that of Bali~et~al. within uncertainties. For order $n=26$ the series of subleading asymptotic terms in the curly brackets in this case read $\{1~(\mbox{NLO}) - 0.725614~(\mbox{NNLO}) - 0.156658~(\mbox{NNNLO})+{\cal O}(1/n^3)\}$, which leads to a quite different outcome.\footnote{In the common $\overline{\rm MS}$ scheme for the strong coupling the corresponding series at order $n=26$ reads $\{1~(\mbox{NLO}) +0.11329~(\mbox{NNLO}) +0.0143241~(\mbox{NNLO})+{\cal O}(1/n^3)\}$, which is very close to the inverse of the unexpanded result given in footnote~\ref{ftnt:MSbar}.}
The results of this approach to determine ${\cal N}_P$ are displayed in the right panel of Fig.~\ref{fig:pinedaplots}. We see that the NNLO and NNNLO results are significantly different than those given in the left panel. Following again Bali~et~al.\ and using the difference between the NNLO and NNNLO results at $n=26$ as the uncertainty, we obtain ${\cal N}_P=(12\pm 16)\times 10^4$, which is compatible with zero. This result corresponds to $N_g^{(n_f=0)} = 4.66 \pm 6.22$ for the GC renormalon normalization in our convention which is perfectly consistent with Eq.~(\ref{eq:Ngoptsubnf0}) and the result of Lee~\cite{Lee:2011te}. The result is, however, only marginally compatible with the result $N_g^{(n_f=0)} = 16.43 \pm 6.73$ based on the unexpanded ratio $p_n^{\rm latt}/p_n^{\rm latt(asy)}$, showing that the uncertainty estimate based on the difference between the NNLO and NNNLO results may not be quite reliable.

Overall, we conclude that the lattice method to determine the GC renormalon norm advocated by Bali et al.\ should be assigned a larger theoretical uncertainty than that quoted in Ref.~\cite{Bali:2014fea} due to the large errors in quantifying the large-order asymptotic behavior of lattice perturbation theory arising from the GC renormalon. This uncertainty stems from the enormous size of the 3- and 4-loop coefficients of the QCD $\beta$-function in the lattice UV renormalization scheme.

\end{appendix}

\bibliography{./sources}
\bibliographystyle{JHEP}

\end{document}